\documentclass[twocolumn,amsmath,superscriptaddress,amssymb]{revtex4}

\begin{document}

\def\beqa{\begin{eqnarray}}
\def\eeqa{\end{eqnarray}}
\def\beqn{\begin{equation}}
\def\eeqn{\end{equation}}

\def\g{g}
\def\R{R}                 
\def\RE{E}                  
\def\REs{\RE^0}
\def\REt{\RE^1}  
\def\REa{\RE^a}  
\def\PN{\Phi_N}
\def\PNn{\Phi_N^{(1)}}
\def\PNN{\Phi_N^{(2)}}
\def\PNa{{\Phi_a}}
\def\PP{{\Phi_P}}
\def\Pst{\phi}
\def\PNan{\delta \PN}
\def\h{h}                 
\def\hht{h_{00}}
\def\htx{h_{0r}}
\def\hhx{h_{11}}
\def\hhr{h_{rr}}
\def\Ps{\Phi^0}
\def\Pt{\Phi^1}
\def\Pa{\Phi^a}
\def\D{D}                 
\def\Gxx{{\Gamma^-_{\s\s}}}   
\def\Gpxx{{\Gamma^+_{\s\s}}}
\def\Gyy{{\Gamma_{\sy\sy}}}
\def\Gst{[\Gamma]_{st}}
\def\Gan{\delta \Gamma}

\def\td{{\cal T}}         
\def\tdxy{{\cal T}_{12}}         
\def\Td{\Delta \td}             
\def\Tdp{{\Delta \td^\prime}}
\def\TdP{{\Delta \td_P}}
\def\TdpP{{\Delta \td^\prime_P}}
\def\tdst{[\td]_{st}}
\def\tdan{\delta \td}

\def\k{k}                  
\def\ks{\mathbf{\k}}
\def\n{n}                  
\def\nxm{\n_1^-}
\def\nxp{\n_1^+}
\def\nym{\n_2^-}
\def\nyp{\n_2^+}
\def\nz{{\bar \n}}
\def\nzxm{{\bar \n}_1^{-}}
\def\nzxp{{\bar \n}_1^{+}}
\def\nzym{{\bar \n}_2^{-}}
\def\nzyp{{\bar \n}_2^{+}}
\def\nzpm{{\bar \n}^{\pm}}
\def\x{x}                  
\def\xx{\x_1}
\def\xxm{\x_1^-}
\def\xxp{\x_1^+}
\def\xy{\x_2}
\def\t{t}                  
\def\tx{\t_1}
\def\txm{\t_1^-}
\def\txp{\t_1^+}
\def\tpx{\t^\prime_1}
\def\ty{\t_2}
\def\xb{\mathbf{\x}}       
\def\xbx{\xb_1}
\def\xbxm{{\xb_1^-}}
\def\xbxp{{\xb_1^+}}
\def\xby{{\xb_2}}
\def\partialx{\partial_{\bold{1}}} 
\def\partialy{\partial_{\bold{2}}}
\def\Dx{{\Delta \x}}
\def\Partial{\nabla}       
\def\s{s}                  
\def\sp{{\s^\prime}}
\def\sxm{\s_1^-}
\def\sxp{\s_1^+}
\def\sl{\sigma}
\def\sx{\s_1}
\def\spx{\s^\prime_1}
\def\sy{\s_2}      
\def\u{u}               
\def\ux{\u_1}
\def\uy{\u_2}  
\def\uxm{\u_1^-}
\def\uxp{\u_1^+}
\def\az{{a_{0}}}           
\def\a{a}
\def\ast{[\a]_{st}}
\def\aan{\delta \a}
\def\aant{\delta \a_\td}
\def\aang{\delta \a_\Gamma}
\def\r{r}                  
\def\rx{{\r_1}}
\def\rxy{{\r_{12}}}
\def\ry{{\r_2}}
\def\rhx{\rho_1}
\def\rhy{\rho_2}
\def\rpy{{\r^\prime_2}}
\def\rpxy{{\r^\prime_{12}}}
\def\rpy{{\r^\prime_2}}
\def\rL{{\r_\Lambda}}
\def\d{d}
\def\ri{\rho}              
\def\th{\theta}            
\def\thx{{\th_1}}
\def\thy{{\th_2}}
\def\Th{\Theta}            
\def\Thst{[\Th]_{st}}        
\def\Than{\delta \Th}        
\def\Thz{\Theta_0}
\def\dTh{{\Delta \Th}}
\def\Thr{{\theta_r}}
\def\Ths{{\theta_s}}
\def\Thx{{\chi}}
\def\ang{\varphi}
\def\angx{{\ang_1}}
\def\dangx{{\dot \ang}_1}
\def\angy{{\ang_2}}
\def\angi{{\ang_0}}
\def\angpx{{\angp_1}}
\def\angpy{{\angp_2}}
\def\thn{\psi}
\def\thnz{\thn_0}
\def\thnst{[\thn]_{st}}      
\def\thnan{\delta \thn}      
\def\thnx{{\thn_1}}
\def\thny{{\thn_2}}
\def\vang{v_\ang}      
\def\vr{v_\r}

\def\S{S}
\def\Sr{{\S_\r}}
\def\Srn{\Sr^{(0)}}
\def\wt{w}              
\def\sw{{s_\wt}}
\def\cw{{c_\wt}}
\def\dw{{d_\wt}}

\def\nux{{\nu_1}}
\def\nuxm{\nu^-}
\def\nuxp{\nu^+}          
\def\nupx{{\nu^\prime_1}}
\def\nuy{{\nu}}
\def\ynu{y}
\def\ynup{y_+}
\def\ynuan{\delta \ynu}

\def\xg{\zeta}
\def\xgp{{\xi}}

\def\T{T}
\def\TTt{{\bar \rho}}

\def\pp{\pi}
\def\ps{\pi^{0}}
\def\pt{\pi^{1}}
\def\pl{\pi^{2}}

\def\M{M}                  
\def\m{m}                  
\def\Gs{G^0}               
\def\Gt{G^1}
\def\Ga{G^a}
\def\Gm{G_m}
\def\GN{G_N}
\def\GNa{G_a}
\def\GP{G_P}
\def\Gta{{\tilde G^a}}
\def\Gts{{\tilde G^0}}
\def\Gtt{{\tilde G^1}}
\def\GtN{{\tilde G_N}}
\def\GtNa{{\tilde G_a}}
\def\GtP{{\tilde G_P}}       
\def\cs{\zeta^0}          
\def\ct{\zeta^1}
\def\ca{\zeta^a}
\def\cN{\zeta_N}
\def\cNa{\zeta_a}
\def\cP{\zeta_P}
\def\chs{\chi^0}         
\def\cht{\chi^1}
\def\cha{\chi^a}

\def\c{c}                  
\def\MS{M}                 
\def\rS{{\rho_S}}
\def\thnS{{\thn_S}}
\def\MT{{M_E}}       
\def\dT{{r_E}}
\def\rT{{r_E}}
\def\rM{r_{M}}             
\def\Ec{{E_c}}             
\def\vP{{v_P}}
\def\dP{{r_P}}

\def\E{E}                  
\def\J{J}                  
\def\K{K}                    

\def\stand#1{\left[#1\right]_\mathrm{st}}

\def\eprint#1{{\it Preprint} #1}

\title{Post-Einsteinian tests of linearized gravitation }

\author{ Marc-Thierry  Jaekel}

\affiliation{Laboratoire de Physique Th\'eorique de l'ENS,
24 rue Lhomond, F75231 Paris Cedex 05 \footnote{
Centre National de la Recherche Scientifique (CNRS), Ecole Normale Sup\'{e}rieur
e (ENS), Universit\'{e} Pierre et
Marie Curie (UPMC); email:jaekel@lpt.ens.fr}}

\author{Serge Reynaud }

\affiliation{Laboratoire Kastler Brossel, case 74, Campus Jussieu,
F75252 Paris Cedex 05 \footnote{CNRS, ENS, UPMC;
email:reynaud@spectro.jussieu.fr}}

\begin{abstract}
The general relativistic treatment of gravitation can be extended
by preserving the geometrical nature of the theory but modifying 
the form of the coupling between curvature and stress tensors. 
The gravitation constant is thus replaced by two running 
coupling constants which depend on scale and differ in the sectors 
of traceless and traced tensors. 
When calculated in the solar system in a linearized approximation,
the metric is described by two gravitation potentials.
This extends the parametrized post-Newtonian (PPN) phenomenological framework
 while allowing one to preserve 
compatibility with gravity tests performed in the solar system.
Consequences of this extension are drawn here for phenomena 
correctly treated in the linear approximation.
We obtain a Pioneer-like anomaly for probes with an eccentric motion 
as well as a range dependence of Eddington parameter 
$\gamma$ to be seen in light deflection experiments. 


\end{abstract}

\maketitle

\section{Introduction}

Experimental tests of gravity performed in the solar system 
show a good agreement with General Relativity.
This statement can be put under a quantitative form by using the
parametrized post-Newtonian (PPN) formalism \cite{Will}
or by bounding deviations of the gravity force law from its
standard form \cite{Fischbach}.
General Relativity is however challenged by observations performed 
at galactic or cosmological scales. 
Due to the good agreement of tests with General Relativity, 
the anomalies seen in the rotation curves of galaxies or in the relation
between redshifts and luminosities for supernovae are commonly accounted
for by introducing dark matter and dark energy components designed to this
purpose \cite{Aguirre,Riess,Perlmutter}. 
As long as these new components are not detected by other means, 
the anomalies can also be thought of as consequences of modifications 
of gravity laws at large scales \cite{Sanders02,Lue04,Turner04}.

An important requirement to be met by such modifications is
that they remain compatible with the bounds set by solar system tests.
The recently observed Pioneer anomaly might be a key piece of information in 
this context by pointing at some anomalous behaviour of gravity at a scale 
of the order of the size of the solar system \cite{Anderson98}. 
The effect appears as an anomalous acceleration recorded on Pioneer 10/11 
probes during their flight to the outer solar system.
The measured Doppler tracking data are compared with predictions of
General Relativity and the residuals then expressed as an unmodelled 
acceleration for the probes.
This anomalous acceleration is directed towards the Sun with an approximately 
constant amplitude $8\times 10^{-10} {\rm{m s^{-2}}}$ over a large range,
20 to 70 astronomical units (UA), of heliocentric distances \cite{Anderson02}.

Up to now, the anomaly has escaped all attempts of explanation as a systematic 
effect generated  by the spacecraft itself or its environment, 
and it has no more been convincingly derived as a consequence of new physics 
(see the references in \cite{Anderson03,Nieto04}).
This status should motivate further scrutiny of any potential origin of the anomaly. 
The aim of a better understanding of deep space navigation is by itself a 
sufficient motivation \cite{Turyshev04} and the possibility that the Pioneer anomaly 
be the first hint of a modification of gravity laws at large scales cannot be 
let aside investigations \cite{Bertolami04}.
  
The long standing opposition between General Relativity and Quantum Theory 
has led to the view that the former can only be an effective theory of 
gravity which could be altered at short or large length scales. 
Radiative corrections naturally lead to an immersion of General Relativity 
within a class of fourth order theories \cite{deWitt,Deser74,Capper74}
with potential effects at short ranges \cite{Stelle}.
Modifications may also be expected to appear at larger length scales
\cite{Sakharov,Adler} with implications in astrophysics \cite{Nieto} 
and cosmology \cite{Deffayet02,Dvali03,Gabadadze04}.
In contrast to Einstein theory which is natively non renormalizable \cite{tHooft}, 
fourth order theories show renormalizability as well as 
asymptotical freedom at high energies \cite{Stelle,Fradkin}. 
The extension of gravitation theory at scales not already constrained by experiments 
can thus be tackled by studying renormalization group trajectories \cite{Reuter}. 

Renormalizability of the family of fourth order theories however comes with a 
counterpart, namely the problem of unitarity associated with ghosts which
is often thought to flaw fundamental field theories.
It has however been convincingly argued that it does not constitute a definitive 
deadend for an effective field theory valid in a limited scale domain \cite{Simon90}. 
In particular, the departure from unitarity is expected to be negligible at 
ordinary scales tested in present day universe \cite{Hawking02}, 
leading to sensible calculations for observable phenomena 
\cite{DeFilippo02,AndersonP03,Hu04}.

In the present paper, we investigate the potential 
consequences of modifying the gravitation theory 
at length scales of the order of the size of the solar system \cite{Jaekel04a}.  
We show that there exist natural extensions of General Relativity 
which preserve its essential geometric features 
such as the identification of gravitation fields with the space-time metric, 
the definition of geodetic motions and the equivalence principle. 
Only the form of the coupling between the Einstein curvature tensor and the 
energy-momentum tensor is modified.
In agreement with behaviours observed for radiative corrections of 
General Relativity, we assume that this coupling can become scale dependent 
and differ in the two sectors of traceless and traced components. 
As the application of the new framework is primarily focused on potential 
consequences in the outer solar system, we then treat the gravitation theory 
in a linearized approximation and describe the gravity source 
as punctual, isotropic and static. 

We thus show that the metric is characterized by two potentials directly related 
to the two running coupling constants which take the place of the single
Newtonian gravitation constant.
The first potential $\PN$ generalizes the usual Newton gravitation potential
and its deviation from the latter suffers stringent bounds deduced from 
the remarkable agreement of General Relativity with gravity tests.
Meanwhile, the second potential $\PP$ opens additional freedom offering 
the possibility to accomodate a Pioneer-like anomaly for probes with 
eccentric trajectories. 
The second potential $\PP$ may also be thought of as promoting the Eddington 
parameter $\gamma$ to the status of a range dependent function.
It thus leads to the prediction of other anomalies to be recorded on null geodesics, 
in particular in time delay, deflection or Doppler tracking experiments performed 
on probes passing behind the Sun 
\cite{Shapiro66,Shapiro71,Reasenberg79,Shapiro99,Bertotti03,Shapiro04}.
These anomalies which have the same origin as the Pioneer-like anomaly in the new 
framework will be discussed in detail below. 
Hopefully, their confrontation to observations will either provide us with data 
points to be compared with the Pioneer data points or lead to new bounds on 
deviations of alternative gravitation theories from General Relativity.
Note that the effects associated with non linearities of gravitation, 
that is also with the Eddington parameter $\beta$, will only be discussed briefly 
since they need not be considered in situations studied here
(more details on this point below).

The following sections begin by a description of the proposed extension 
of Einstein gravitation theory in a linearized approximation.
Considering a punctual, isotropic and static gravitational source, 
we deduce the expression of the modified metric in terms of
two potentials $\PN$ and $\PP$.
We then examine the consequences of the modifications on the third 
Kepler law, on Doppler tracking of probes and on light deflection. 
We finally use the simplest example, which involves only three 
additional parameters, as a benchmark for comparing the predictions 
of the modified framework to available observations.

\section{Modified gravitation equations}
\label{sect:gravitation_equations}

We first introduce the extension of Einstein theory by discussing its motivations 
and writing down modified post-Einsteinian gravitation equations. 
The term `post-Einsteinian' has a twofold meaning alluded to in the Introduction. 
It first means that the geometrical nature of gravity, the very core of Einstein 
theory and, furthermore, one of the best ever tested properties of the physical 
world, is left untouched.
But it also points to the fact that the coupling between curvature
and stress tensors suffers a modification from its Einsteinian form.

As a first step in this discussion, we emphasize that the geometric foundations of 
the theory are preserved~: the gravitational field is identified with a metric 
tensor $\g_{\mu\nu}$, the motions are described by associated geodesics and,
as a consequence, the equivalence principle is left untouched.
As is well known, this principle is verified at distances ranging from the 
millimeter in laboratory experiments (\cite{Adelberger03} and references in) 
to the sizes of Earth-Moon orbit \cite{Williams96} 
or Sun-Mars orbit \cite{Hellings83,Anderson96}.
The relative accuracy of these tests, better than $10^{-12}$ for some of them, 
is good enough to discard any interpretation of the Pioneer anomaly from a 
violation of the equivalence principle \cite{Anderson02}.
This statement does not entail that the equivalence principle is exact,
but it means that the potential deviations from General Relativity studied 
here have a larger amplitude than the violations of the equivalence 
principle expected from unification models \cite{Damour,Overduin00}.
This is why they can be analyzed in a metric theory sharing the conceptual 
basis of General Relativity.

As already discussed, we focus the attention on the linearized theory
where the gravitation field is represented as a small perturbation 
$\h_{\mu\nu}$ of the Minkowski metric $\eta_{\mu\nu}$ 
\beqa
&&\g_{\mu\nu} = \eta_{\mu\nu} + \h_{\mu\nu} \\
&&\eta_{\mu\nu} = {\rm diag}(1, -1, -1, -1) 
\quad,\quad \left\vert \h_{\mu\nu} \right\vert \ll 1 \nonumber
\eeqa
The field $\h_{\mu\nu}$ is written as a function of position $x$
in spacetime or, equivalently, of wavevector $k$
\beqa
\h_{\mu\nu}(\x) \equiv \int {d^4\k \over (2\pi)^4}e^{-i\k\x} \h_{\mu\nu}[\k]
\eeqa 
Riemann, Ricci and scalar curvatures are easily written in momentum representation, 
at first order in $\h_{\mu\nu}$, 
\beqa
\label{curvatures}
&&\R_{\lambda\mu\nu\rho} = \frac{
\k_\lambda\k_\nu \h_{\mu\rho} - \k_\lambda\k_\rho \h_{\mu\nu}
- \k_\mu\k_\nu \h_{\lambda\rho} + \k_\mu\k_\rho \h_{\lambda\nu}}2\nonumber\\
&&\R_{\mu\rho} = \frac{\k^2 \h_{\mu\rho} - \k_\mu\k^\sigma \h_{\sigma\rho}
-  \k_\rho\k^\sigma \h_{\sigma\mu} + \k_\mu\k_\rho \eta^{\mu\nu}\h_{\mu\nu}}2\nonumber\\
&&\R = \k^2 \eta^{\mu\nu}\h_{\mu\nu} -  \k^\mu\k^\nu \h_{\mu\nu}
\eeqa
We use the same sign conventions as in \cite{Landau}, indices being raised or 
lowered using Minkowski metric $\eta_{\mu\nu}$. 

These expressions can be recovered from Einstein curvature tensor $\RE_{\mu\nu}$
\beqa
\RE_{\mu\nu} &\equiv& \R_{\mu\nu} - {1\over 2} \eta_{\mu\nu} \R \\
\R_{\mu\nu} &=& \RE_{\mu\nu} - {1\over 2} \eta_{\mu\nu} \RE 
\quad , \quad \RE = -\R \nonumber\\
\R_{\lambda\mu\nu\rho} &=& \frac{
\k_\lambda\k_\nu \R_{\mu\rho} - \k_\lambda\k_\rho \R_{\mu\nu}
- \k_\mu\k_\nu \R_{\lambda\rho} + \k_\mu\k_\rho \R_{\lambda\nu}}{k^2}\nonumber
\eeqa
$\RE_{\mu\nu}$ may be written in terms of transverse projectors
\beqa
&&\RE_{\mu\nu} = {\pp_{\mu\rho} \pp_{\nu\sigma} + \pp_{\mu\sigma} \pp_{\nu\rho} 
-\pp_{\mu\nu} \pp_{\rho\sigma} \over2} \k^2\h^{\rho\sigma} \\
&&\pp_{\mu\nu}  \equiv \eta_{\mu\nu} - {\k_\mu\k_\nu \over \k^2} \quad , \quad 
\k^2 \equiv \k^\mu \k_\mu \quad , \quad \k^\mu \pp_{\mu\nu}=0 \nonumber\\
&&\pp_{\mu\nu}{\pp^\nu}_\rho = \pp_{\mu\rho} \quad , \quad 
\pp_{\mu\nu} \pp^{\mu\nu} = 3\nonumber
\eeqa
It may then be decomposed as the sum of two independent components with 
different conformal weights \cite{Jaekel95}
\beqa
\label{Einstein_curvatures}
\RE_{\mu\nu} &\equiv&\REs_{\mu\nu} + \REt_{\mu\nu}\nonumber\\
\REs_{\mu\nu} &\equiv& \ps_{\mu\nu\rho\sigma} \RE^{\rho\sigma}
=  {1\over2} \ps_{\mu\nu\rho\sigma}
\k^2\h^{\rho\sigma} \nonumber\\
\REt_{\mu\nu} &\equiv& \pt_{\mu\nu\rho\sigma} \RE^{\rho\sigma}
=- \pt_{\mu\nu\rho\sigma} \k^2 \h^{\rho\sigma}
\eeqa
where $\ps$ and $\pt$ project transverse tensors onto their traceless and 
traced components 
\beqa
\label{projectors}
&&\ps_{\mu\nu\rho\sigma} \equiv 
{\pp_{\mu\rho} \pp_{\nu\sigma} + \pp_{\mu\sigma} \pp_{\nu\rho} \over 2} 
 - \pt_{\mu\nu\rho\sigma} \nonumber\\
&&\pt_{\mu\nu\rho\sigma} \equiv {\pp_{\mu\nu} \pp_{\rho\sigma} \over 3} \\
&&\ps_{\mu\nu\rho\sigma} \pp^{\rho\sigma} = 0 \quad , \quad  
\pt_{\mu\nu\rho\sigma} \pp^{\rho\sigma} =\pp_{\mu\nu} \nonumber
\eeqa

As discussed in the introduction, we disregard the effects of rotation and 
non sphericity of the Sun which affect significantly the motions of inner planets 
but have small effects on motions in the outer solar system or 
on propagation of light. More precisely, we consider that these
effects are properly taken into account in the standard description
based on General Relativity (see for example the calculation of Doppler 
tracking data for Pioneer probes in \cite{Anderson02}) and 
that they have the same impact in the modified and standard descriptions. 
As a consequence, the anomalies evaluated by subtracting the standard result 
from the modified one can be calculated with the assumptions 
of stationarity and isotropy.
The same statements can be applied essentially unmodified to the assumption 
of linearity (see more details on this discussion in the next section).

Fields are written as functions of spherical coordinates 
\beqa
\label{polar_coordinates}
\x^\mu &\equiv& (\c \t, \r \cos\th \cos \ang, 
\r\cos \th \sin \ang, \r\sin \th)
\eeqa
where $\c$ denotes light velocity, $\t$ and $\r$ the time and radius,
$\th$ and $\ang$ the colatitude and azimuth angles.
Isotropic quantities are functions of $\x^0$ and $\r$ only.
Then, the Einstein tensor (\ref{Einstein_curvatures}) 
is read in terms of two potentials
\beqa
\label{isotropic_Einstein}
&&\RE_{\mu\nu} = \ps_{\mu\nu 0 0}\ 2\k^2 \Ps + \pt_{\mu\nu 0 0}\ 2\k^2 \Pt
\eeqa
$\Ps$ and $\Pt$ are functions of $\x^0$ and $\r$ or, alternatively
in momentum representation, of $\k_0$ and $\vert\ks\vert$ with $\k^2=\k_0^2-\ks^2$. 
Conversely, $\Ps$ and $\Pt$ can be written in terms of the traceless and traced 
curvatures (\ref{Einstein_curvatures}) 
\beqa
\label{Eddington_potentials}
2\Ps = {3 \over 2} {\k^2 \over (\ks^2)^2} \REs_{00} 
\quad,\quad
2\Pt = 3 {\k^2 \over (\ks^2)^2} \REt_{00} &&
\eeqa
When stationarity is also assumed, these relations are simplified
due to the fact that $\k_0=0$ and $\k^2=-\ks^2$. 

The Einstein tensor $\RE_{\mu\nu}$ is transverse as a consequence of Bianchi identity
and this is also the case for the stress tensor $\T_{\mu\nu}$, due to energy-momentum 
conservation. The second condition also implies geodesic motion for test masses,
in consistency with the geometrical interpretation of gravity which is preserved  
by the extension of Einsteinian theory studied here.
In General Relativity, these two tensors are merely proportional to each other 
\cite{Landau}
\beqa
\label{GR_gravitation_law}
\RE_{\mu\nu} = 8 \pi \GN \T_{\mu\nu}
\eeqa
with $\GN$ the Newton constant. Now this equation may easily be generalized 
while preserving transversality so that, in the linearized theory, the
relation between the two tensors takes the form 
\beqa
\label{general_gravitation_law}
\REa_{\mu\nu} = \cha_{\mu\nu\lambda\rho} \T^{\lambda\rho}
\quad,\quad a=0,1 &&
\eeqa
The response functions $\cha_{\mu\nu}$ are constrained by the transversality 
condition but they may depend on momentum and differ in the two sectors,
thus extending the form of the coupling assumed in Einsteinian theory. 

We treat the Sun as a static gravitational mass $M$ localized at 
the origin $\x=0$ of the coordinate system
\beqa
\label{point_stress_tensor}
&&\T_{\mu\nu}(\x) = \eta_{\mu0}\eta_{\nu0} \M \c^2\delta^{(3)}(x)\nonumber\\
&&\T_{\mu\nu}[\ks] = \eta_{\mu0}\eta_{\nu0} \M \c^2
\eeqa
We have introduced a notation for the Fourier transform of stationary quantities
\beqa
&&\T_{\mu\nu}[\k] \equiv \T_{\mu\nu}[\ks] 2\pi\delta(\k_0)
\eeqa
The potentials thus have stationary and isotropic expressions 
given by Poisson like equations
\beqa
\label{inv_grav_equations}
-\ks^2 \Pa[\ks] =  {2\Gta[\ks] \M \over \c^2} 
\quad,\quad a=0,1 &&
\eeqa

With the assumptions of stationarity and isotropy, and for a pressureless 
gravitational source (see eq.(\ref{point_stress_tensor})),
the two running constants $\Gta$ constitute the information on the linear 
response functions $\cha$ which can be extracted from experiments
\beqa
\label{general_running_constants}
&&{\Gts[\ks] \over \c^4} \equiv {3 \over 8}  \chs_{0000}[0,\ks] 
\nonumber\\
&&{\Gtt[\ks] \over \c^4} \equiv  {\eta^{\mu\nu}\over 4}
\cht_{\mu\nu 0 0}[0,\ks] 
\eeqa
Einstein theory corresponds to the simple case where the 
two running couplings reduce to a single constant $\GN$.
Note also that these two running constants $\Gta$ do not exhaust 
the information contained in the linear response functions $\cha$.
It is for example easy to show that equation (\ref{general_gravitation_law})
allows for three independent scalar functions while still preserving
transversality, stationarity and isotropy. But the third one cannot be
tested in the solar system due to the absence of pressure terms in the
gravitational source (\ref{point_stress_tensor}).

In order to underline the potential deviations from General Relativity, 
we will introduce a generic notation showing the separation between 
a standard expression $\stand{\ }$ (\textit{ie} the prediction of 
General Relativity) and an anomaly indicated by the symbol $\delta$.
For example, the two potentials are the sums of a standard expression
\beqa
\stand{\Pt(\r)} = \stand{\Ps(\r)} = 
-{\GN \M \over \c^2 \r} \equiv \phi (\r) &&
\eeqa
and of a deviation which has to remain small in order to fit existing gravity tests
\beqa
\label{standard_plus_anomaly}
\Pa (\r) \equiv \stand{\Pa(\r)} + \delta\Pa (\r) \quad,\quad 
\left| \delta\Pa (\r) \right| \ll 1 &&
\eeqa
The symbol $\ll$ is used here to suggest that the quantities $\delta\Pa$
are smaller than the Newtonian potential $\phi$ which is already smaller 
than unity in the linearized approach. Since these functions may have different
dependences versus $\r$, this can only be a qualitative statement at the moment.
The main objective of the present paper is to discuss the potential 
phenomenological consequences of $\delta\Pa$ in the range of 
momentum or, equivalently, of length scales tested in the solar system.

With this purpose in mind, we will avoid favoring any particular functional 
dependence of the anomalous running constants $\delta\Gta$ or of the
associated anomalous potentials $\delta\Pa$.
In order to illustrate the discussion, we will however use occasionally 
as a ``benchmark'' a specific form of the potentials 
\beqa
\label{Newton_Prime_constants}
\Pa(\r) \simeq -{\Ga \M \over \c^2 \r} + {\ca \M \r\over \c^2}
\quad,\quad a=0,1 &&
\eeqa
This form superimposes linear terms proportional to $\r$ to the standard
terms proportional to $1/\r$. 
Equivalently, the running constants are the sums of standard constant
terms and of infrared corrections proportional to $1/\ks^2$
\beqa
\label{Running_constants}
\Gta[\ks] \simeq \Ga + {2\ca \over \ks^2}
\quad,\quad a=0,1 &&
\eeqa
The main advantage of this simple example is that the phenomenology is 
now determined  by four constants, the interpretation of which is given below.
In fact constants of integration appear when solving equations 
(\ref{inv_grav_equations}) with the running constants (\ref{Running_constants}).
They give rise to terms proportional to $\r^0$ and $\r^2$ besides those
written in (\ref{Newton_Prime_constants}).
Since constant corrections to the metric do not affect curvatures
while quadratic terms correspond to constant curvatures
relevant at larger galactic or cosmological scales, 
both can be ignored in the solar system. 

The linear dependence of the terms proportional to $\ca$ 
in (\ref{Newton_Prime_constants}) also calls for remarks. 
Should it hold over galactic length scales, this dependence would induce 
effects incompatible with observations. 
This means that (\ref{Newton_Prime_constants}) can only be considered 
as effective potentials valid in restricted ranges of distances or momenta.
The linear terms can for example result from an expansion at $\r\ll\rL$
of Yukawa type potentials $(1/r)\exp(\r/\rL)$ which are 
regular at large distances \cite{Jaekel04b}.
Note that radiative corrections to General Relativity are described by 
modifications of the lagrangian which behave, at lowest order, as quadratic 
forms of the curvature tensors \cite{deWitt,Deser74,Stelle,Sakharov} and 
then give rise, in the linear approximation, to Yukawa corrections.
Though it is not possible to meet the goal of accomodating the Pioneer anomaly 
in a frame compatible with planetary tests by adding Yukawa corrections to 
the single Newton potential \cite{Jaekel04b}, the presence of two potentials 
with different values in the two sectors now opens an additional freedom.

In the following, we consider modifications of the gravity theory described
by two potentials $\Pa(\r)$. Occasionally, we use the simple forms 
(\ref{Newton_Prime_constants}) or (\ref{Running_constants}) 
for deriving preliminary predictions for experiments in the solar system,
keeping in mind that a large distance regulator is implicit when discussing for 
example the deflection of light coming from stars or extragalactic sources.

\section{Modified metric tensors}
\label{sect:metric_tensors}

We now write the metric obtained as a solution of the modified equations 
(\ref{inv_grav_equations},\ref{general_running_constants})
and to be used for discussing gravity tests in
the forthcoming sections. 

To this aim, we adopt the gauge convention of the PPN formalism \cite{Will} 
with the metric written as a stationary and isotropic function of spherical 
coordinates (\ref{polar_coordinates})
\beqn
\label{isotropic_metric}
d\s^2 = \g_{00} \c^2 d\t^2 + \g_{\r\r} \left( d\r^2  +
\r^2(d\th^2 + {\rm \sin}^2\th  d\ang^2) \right)
\eeqn
This gauge choice fixes the expressions of the metric components
in terms of the two potentials (\ref{Eddington_potentials}) 
\beqa
\label{polar_metric_NP}
\g_{00} &=& 1 + 2 \PN \quad,\quad \PN \equiv {4 \Ps - \Pt \over 3} \\
\g_{\r\r} &=& -1 + 2 \PN - 2\PP \quad,\quad 
\PP \equiv {2 (\Ps - \Pt) \over3} \nonumber
\eeqa
$\PN$ and $\PP$ have been introduced as two linear combinations of $\Ps$ and $\Pt$. 
$\PN$ is defined from the difference $(\g_{00}-1)$ and thus
identified as a generalization of Newton potential $\Pst$
while $\PP$ is defined from $(-\g_{00} \g_{\r\r}-1)$ and
measures the difference between the two 
sectors of traceless and traced curvatures.

As $\Ps$ and $\Pt$, $\PN$ and $\PP$ obey Poisson equations
with running constants $\GtN$ and $\GtP$ written as
linear combinations of $\Gts$ and $\Gtt$ 
\beqa
&&-\ks^2 \PNa[\ks] =  {2\GtNa[\ks] \M \over \c^2} 
\quad,\quad a= N,P  \\
&&\GtN \equiv {4 \Gts - \Gtt \over 3} \quad,\quad 
\GtP \equiv {2 (\Gts - \Gtt) \over3}  \nonumber
\eeqa
Einstein theory corresponds to the standard expressions
\beqa
\stand{\PN(\r)} \equiv \phi (\r) = -{\GN \M \over \c^2 \r} \quad,\quad 
\stand{\PP(\r)} = 0 &&
\eeqa
and deviations have to remain small
\beqa
\label{two_potentials}
\PNa (\r) \equiv \stand{\PNa(\r)} + \delta\PNa (\r) \quad,\quad 
\left| \delta\PNa (\r) \right| \ll 1 &&
\eeqa

According to the discussion of the preceding section, we will occasionally use 
the specific form (\ref{Newton_Prime_constants}) of the potentials 
\beqa
\label{Newton_Prime_constants_NP}
\PNa (\r) = -{\GNa \M \over \c^2 \r} + {\cNa \M \r\over \c^2} \quad,\quad
a=N,P &&
\eeqa
with the new coeffients obtained as in (\ref{Running_constants}).
With the simple model (\ref{Newton_Prime_constants_NP}), the phenomenology 
is determined  by four constants, the Newton constant $\GN$ and three small 
parameters $\GP$, $\cN$, $\cP$. 
This allows one to reach more specific conclusions with however
the drawback of a loss of generality with respect to (\ref{two_potentials}).
This is why we will as far as possible use the general form 
(\ref{polar_metric_NP}) of the metric and restrict the use of the simple
model (\ref{Newton_Prime_constants_NP}) to a few specific occasions.

At this point, it is worth comparing the metric (\ref{polar_metric_NP}) with 
the parametrized post-Newtonian (PPN) expression 
introduced by Eddington \cite{Eddington} and then developed by  
several physicists \cite{Robertson,Schiff66,Nordtvedt68,WillNordtvedt72}. 
The Eddington PPN metric is read in the isotropic gauge (\ref{isotropic_metric}) 
\beqa
\label{PPN_0}
&&\g_{00} = 1 + 2 \alpha \phi + 2 \beta \phi^2 + \ldots \nonumber \\ 
&&\g_{\r\r} = -1 + 2 \gamma \phi + \ldots 
\eeqa
As the three Eddington parameters are constant, the first one $\alpha$ is usually 
eliminated through a redefinition of the Newton constant.
We show now that the metric (\ref{polar_metric_NP}) can be regarded as similar to 
(\ref{PPN_0}) but with the quantities $\alpha$, $\beta$ and $\gamma$ 
promoted from the status of a constant to that of a function.
To this aim, we rewrite (\ref{PPN_0}) as the sum of a standard expression obtained for 
$\alpha=\beta=\gamma=1$ and of a deviation $\delta\g_{\mu\nu}$
\beqa
\label{PPN}
&&\delta\g_{00} \simeq  2 (\alpha -1) \phi + 2 (\beta -1) \phi^2 + \ldots \nonumber \\ 
&&\delta\g_{\r\r} \simeq  2 (\gamma -1) \phi + \ldots 
\eeqa
and we identify the terms appearing in (\ref{PPN}) with the anomalous potentials
in (\ref{two_potentials}). 

First, the term $(\alpha -1) \phi $ in (\ref{PPN}) becomes the function $\delta\PN$.
It may have a more general $\r$-dependence which also means that $\alpha$ 
is no longer a constant. 
Similarly, the potential $\PP$ is another function of $\r$ which somewhat
generalizes the term $(\gamma -1)\phi$ in (\ref{PPN}).
As a consequence, gravity tests expressed as an upper bound on the constant 
$(\gamma -1)$ in the PPN formalism will become a bound on the 
function $\PP$ in the new framework.
In loose words, the bound on the Eddington parameter $\gamma$ may now become 
non uniform at different distances in the extended framework.
This statement will be made more quantitative in the following. 
We will see that a long range variation of $\PP$ can lead to a gravitational 
interpretation of the Pioneer anomaly while circumventing the main objection 
to such an interpretation, namely the absence of a similar effect 
on planets \cite{Anderson02}. 
The presence of the second potential $\PP$ at short heliocentric distances 
will also affect time delay or light deflection experiments, 
opening the possibility to see it by reanalyzing data of 
deflection experiments \cite{Bertotti03}.

The manifestation of the second potential $\PP$ in anomalies usually 
ascribed to $(\gamma-1)$ is our primary subject of interest here.
Before focusing the discussion on this subject,
let us open a parenthesis on the non linearities of gravitation theory 
described by the parameter $\beta$ in the PPN metric (\ref{PPN}).
These effects are known to be significant in the inner part of the solar 
system by affecting the perihelion precession of Mercury \cite{Shapiro89} 
as well as the polarization of the orbit of Moon around Earth induced 
by the Sun \cite{effetNordtvedt}.
In contrast, they are small for probes in the outer part of the solar 
system or for propagation of light and this is a first good 
reason for disregarding them in the context of the present paper. 
Let us briefly sketch how this simple argument can be put in a more
quantitative form.

The expansion (\ref{polar_metric_NP}) of $\g_{00}$ can be pushed one order 
further to obtain an expression comparable to the PPN expression (\ref{PPN})
\beqn
\label{second_order_g00}
\delta{\g_{00}} = 2\delta\PNn + 2 \delta\PNN 
\eeqn
$\PNn$ is just the first order potential discussed up to now and 
$\PNN$ the second order contribution to the metric component $\g_{00}$.
Then $\delta\PNN$ is the anomalous part of $\PNN$ which takes the place 
occupied by $(\beta-1)\phi^2$ in the PPN expression (\ref{PPN}).
In loose words, $\delta\PNN$ promotes a constant anomaly $(\beta-1)$ 
to the status of an extra function. 
This function should play a significant role in the evaluation of 
the Mercury perihelion precession or of the orbit of Moon,
a discussion which requires a non linear treatment of gravitational 
perturbations and is not done in the present paper.
However we want to take care of its existence if only to make clear 
that it does not spoil the conclusions of the linear treatment.

To this aim, we now consider an argument which plays a key role in the
presentation of the PPN metric (\ref{PPN}), given up to first order
for $\g_{\r\r}$ but to second order for $\g_{00}$.
The idea is that terms proportional to $\beta$ may have the same impact on 
motions than those proportional to $\gamma$, though they appear at the 
next order in the metric.
The spatial metric components are indeed multiplied by the
kinetic factor $v^2 /c^2$ in the action of test masses and this factor 
has the same magnitude as $\left|\phi\right|$ for virialized motions, 
in particular for circular motions.
Now, as a consequence of the same argument, the effects associated to 
$\gamma$ should dominate those associated to $\beta$ as soon as the kinetic 
factor $v^2 /c^2$ is much larger than $\left|\phi\right|$. 
This is the case for the two situations studied in the present paper, 
namely the motion of Pioneer-like probes escaping the solar system 
and the deflection of light.
In other words, this argument allows one to delineate the category of phenomena 
correctly accounted for in the linearized treatment and the problems 
studied in this paper are precisely chosen within this category.

\section{Modified geodesics}
\label{sect:geodesics}

The geometric foundations of General Relativity are left unchanged
in the new framework and motions are described by geodesics associated 
with the stationary and isotropic metric (\ref{polar_metric_NP}).
We characterize these geodesics and, as a first application, 
we evaluate the modification of the third Kepler law.

The motion of a probe may be written down as 
the Hamilton-Jacobi equation for the action $\S$, that is equivalently
the phase of the associated field, 
\beqa
\label{Hamilton_Jacobi}
&&\g^{\mu\nu} \partial_\mu \S \partial_\nu \S = \m^2 \c^2
\eeqa
$\g^{\mu\nu}$ is the inverse of the metric
and $\m$ the mass of the probe, with $\m \equiv 0$ for light.
This equation is easily solved for a stationary 
and isotropic metric \cite{Landau} with no
dependence on time $t$ and angle variables $\ang$ and $\th$
\beqn
\partial_t \g_{\mu\nu} = \partial_\ang \g_{\mu\nu} = 
\partial_\th \g_{\mu\nu} = 0
\eeqn
Energy $\E$ and angular momenta $\J$ and $\K$ are conserved
\beqa
\label{conservations}
&& \E = -\partial_\t \S = \m \c^3 \g_{00} {d\t \over d\s} \nonumber\\
&& \J = -\partial_\ang \S = \m \c \g_{\ang\ang} {d\ang \over d\s} 
\nonumber\\
&&\K = -\partial_\th \S = \m \c \g_{\th\th} {d \th \over d\s} 
\eeqa
and the action $\S$ can be written in terms of 
a phase shift function $\Sr$ depending only on $\r$ 
\beqa
\label{phase_shift}
&&\S = -\E \t + \J \ang + \Sr \nonumber\\
&&\partial_\r \S = \partial_\r \Sr = -\m\c\g_{\r\r} {d\r \over d\s} 
\eeqa
Without loss of generality, the trajectory has been assumed to take 
place in the plane $\th = {\pi \over2}$ ($\K=0$). 

The Hamilton-Jacobi equations (\ref{Hamilton_Jacobi}) also expresses
the normalization of the 4-velocity $u^\mu$ 
\beqa
\label{velocity_normalization}
u^\mu &\equiv& {d\x^\nu \over d\s} \quad,\quad
g_{\mu\nu} u^\mu u^\nu =1 \\
{\E^2 \over \m^2 \c^4} &=&
\g_{00}\left( 1 - {1\over\g_{\r\r}} {\J^2 \over  \m^2 \c^2 \r^2} \right) 
- \g_{00}\g_{\r\r}\left({d\r \over d\s}\right)^2 \nonumber
\eeqa
With the metric (\ref{polar_metric_NP}), this relation is read as follows 
in terms of the potentials
\beqa
{\E^2 \over \m^2 \c^4} 
&=& 1 + 2 \PN + \left(1 + 4 \Ps\right) {\J^2 \over  \m^2 \c^2 \r^2} \nonumber\\
&& + \left(1 + 2 \PP\right) \left({d\r \over d\s}\right)^2
\eeqa
Note that $\left(1 + 4 \Ps\right)$ is the linearized form of 
the conformally invariant quantity $\left(-\g_{00}/\g_{\r\r}\right)$ 
(see eq.(\ref{polar_metric_NP})).

Another form of the equation of motion is the geodetic equation 
obtained by differentiating (\ref{Hamilton_Jacobi}) or
(\ref{velocity_normalization})
\beqa
\label{geodetic}
&&{\D u_\mu \over d\s} \equiv {d u_\mu \over d\s} 
- \Gamma_{\mu,\nu\rho} u^\nu u^\rho = 0 \\
&&\Gamma_{\mu,\nu\rho} \equiv {1\over2}\left(\partial_\nu\g_{\mu\rho}
+ \partial_\rho\g_{\mu\nu} - \partial_\mu\g_{\nu\rho}\right) \nonumber
\eeqa
$\D$ is the covariant derivative and $\Gamma_{\mu,\nu\rho}$ the 
Christoffel symbols associated with the metric. 
This equation can also be written in terms of accelerations defined 
as Christoffel symbols projected along the motion
\beqa
\label{Christoffel}
{d u_\mu \over d\s} &=& \Gamma_{\mu,\s\s} 
\equiv \Gamma_{\mu,\nu\rho} \u^\nu \u^\rho 
\eeqa
In particular, the radial acceleration can be rewritten 
\beqa
\label{acceleration_field}
\Gamma_{\r,\s\s} &=&
- \left(1 - 2 \PN\right) \left( {d\PN \over d\r} 
+ {d\PP \over d \r}\left({d\r\over ds}\right)^2 \right)\nonumber\\
&&- {1 - 2 \PN \over 2}{d\over d\r}\left(\left(1 + 4 \Ps\right)
{\J^2 \over \m^2\c^2 \r^2}\right) 
\eeqa 
In the following, we use this expression linearized at first order in the 
gravitational perturbation and written as the sum of the standard
expression and of an anomaly  
\beqa
\label{anomalous_acceleration_field}
\Gamma_{\r,\s\s} &=& \stand{\Gamma_{\r,\s\s}} + \delta \Gamma_{\r,\s\s}\nonumber\\
\stand{\Gamma_{\r,\s\s}} &=& - {d\Pst \over d\r}
+\left(1 + 2 \Pst - 2\r {d\Pst \over d\r}\right){\J^2 \over \m^2\c^2 \r^3} \nonumber\\
\delta \Gamma_{\r,\s\s} &=&  - {d \delta \PN \over d\r}
+\left(1 + 2 \delta \PN  - 2\r {d\delta \PN \over d\r}\right){\J^2 \over \m^2\c^2 \r^3}
\nonumber\\
&+&{d\over d\r}\left(\PP {\J^2 \over \m^2\c^2 \r^2}\right) - {d\PP \over d \r}
\left({d\r\over ds}\right)^2
\eeqa
This expression makes it easy to identify the contributions of $\delta\PN$ and $\PP$ 
to potential anomalies to be seen in the weak field approximation. 
The role played by the contribution of $\PP$ in Pioneer like anomaly and deflection
tests is discussed in detail in the next sections.  

As a first application, we now evaluate the modification of the third Kepler law
for circular orbits which depends essentially on $\delta\PN$.
To this purpose, it is convenient to introduce the squares of the radial and angular
velocities with respect to time, $\vr^2$ and $\vang^2$ respectively,
\beqa
\label{squared_velocities}
&&\vr^2 \equiv -\g_{\r\r} {\dot \r}^2 \quad,  \quad
{\dot \r} \equiv {d \r \over d\t}\nonumber\\ 
&&\vang^2 \equiv -\g_{\r\r} \r^2 {\dot \ang}^2 \quad ,
\quad {\dot \ang} \equiv {d\ang \over d\t}
\eeqa
We now differentiate the normalization relation (\ref{velocity_normalization}) 
with the constraints that $\E$ and $\J$ are conserved and obtain a generalized Kepler law 
\beqn
\label{generalized_Kepler}
{\vang^2 \over \c^2} =
{1 \over 2} {\r \over (1 - {1\over 2} \r \partial_\r \g_{\r\r})}
\left( \partial_\r\g_{00}
+ \g_{00}^2 \partial_\r \left({\vr^2\over\g_{00}^2 \c^2}\right)\right) 
\eeqn
For circular orbits  ($\vr =0$), this law is read
\beqn
\label{Kepler}
{\vang^2 \over \c^2 } = -{\r \partial_\r \g_{00} \over
 \g_{\r\r}(2 - \r \partial_\r \g_{\r\r})} 
\eeqn

For weak gravitational fields and at leading order, this Kepler law 
depends only on $\g_{00}$, that is also on $\PN$, 
\beqn
\label{Kepler2}
\vang^2 \simeq \r^2 {\dot \ang}^2 \simeq    
\c^2 \r\partial_\r \PN
\eeqn
An anomaly $\PNan$ of $\PN$ leads to an anomaly 
in the third Kepler law for a circular orbit, as is made evident by 
decomposing $\vang^2$ in standard and anomalous parts
\beqa
\label{third_Kepler_law}
\vang^2 &=& \stand{\vang^2} + \delta \vang^2 \\
\stand{\vang^2} &\simeq& \c^2 \r\partial_\r \Pst \quad , \quad
\delta \vang^2 \simeq \c^2 \r\partial_\r \PNan \nonumber
\eeqa
The agreement between General Relativity and tests performed on planets 
tells us that $\partial_\r \PNan$ has to be small. 
In any case, it is certainly too small to explain the Pioneer anomaly 
(see the discussions in \cite{Anderson02} and \cite{Jaekel04b}). 
Consequently, the deviations from General Relativity studied in the present paper 
are mostly ascribable to the effects of the second potential $\PP$ discussed in
the next sections. 

Note however that more accurate statements taking into account higher order effects 
can be obtained by applying exact expressions written in the present section
to the metric. 
According to the argument presented at the end of section \ref{sect:metric_tensors},
this analysis should let the conclusions of the present paper unaffected.

\section{Doppler tracking}
\label{sect:Pioneer}

We show now that the second potential $\PP$, which does not affect appreciably
the third Kepler law for circular motions, may in contrast lead to 
a Pioneer like anomaly for probes having an eccentric motion. 
To this aim, we perform within the extended 
framework the calculation of Doppler tracking of such probes. 
We do not enter into the technical details described in the analysis 
of the Pioneer anomaly in \cite{Anderson02}.

As a first step, we write the expression of the Doppler frequency shift 
and its time variation in terms of the two gravitation potentials and of 
the probe velocity. 
The tracking is built up on lightlike signal propagating from Earth to the 
remote probe and back from the probe to Earth.
Earth follows a nearly circular geodesic whereas the probe follows an eccentric 
geodesic escaping the solar system.
The up-link radio signal leaves the Earth at the location $\xxm = (\txm, \xbxm)$
and attains the probe at $\xy= (\ty,\xby)$.
There, it is transponded into a down-link radio signal leaving the probe at $\xy$ 
and reaching the Earth at $\xxp=(\txp,\xbxp)$.
The two links are null geodesics with their endpoints, corresponding to emission 
and reception, on the geodesics of the station on Earth and the probe.
They are described by a function $\td$ which measures the time taken by the 
lightlike signal to propagate from the emitter to the receiver
\beqa
\label{lightlike_links}
&&\ty = \txm + \td(\xbxm, \xby) \nonumber\\
&&\txp = \ty + \td(\xbxp, \xby) 
\eeqa
This function $\td$ is calculated by considering the up and down links 
as the advanced and retarded parts of a lightcone originating from $\xy$.
As there is only one lightlike path from a spatial point to another in
the weak gravitational field of the solar system, this definition 
is unambiguous. For the metric (\ref{polar_metric_NP}) furthermore, 
the function $\td$ is symmetric in its two arguments.
Its explicit expression is derived from the action (\ref{phase_shift}) 
in appendix \ref{appendix_time_delay}.

The endpoints $\xx^\pm$ and $\xy$ follow geodesics with 
velocities $\ux^\pm$ and $\uy$.
Since they are related by equations (\ref{lightlike_links}), 
the corresponding proper times $d\sxm$, $d\sxp$ and $d\sy$ obey
\beqn
\label{proper_times}
-{(\uxm \cdot \nxm) \over (\uy \cdot \nym)}d\sxm = d\sy =
-{(\uxp\cdot \nxp)\over (\uy \cdot \nyp)}d\sxp
\eeqn
$\n^\pm_a$ represent the directions of wavevectors at endpoints $a=1,2$
and $\cdot$ denotes a 4-dimensional scalar product
\beqa
\label{directions_waves}
&&(\n^\pm_1)_0 = - (\n^\pm_2)_0 = 1 \quad , \quad
(\n^\pm_a)_i = \mp \c {\partial \td\over \partial (\xb_a^\pm)^i} \nonumber\\
&&(\u \cdot \n) \equiv \u^\mu \n_\mu
\eeqa
The up-link signal leaves the Earth with a frequency determined by 
comparison with a clock located at the emission station, it travels to the 
probe where it is transponded and multiplied by a constant factor.
It travels back to the Earth where the frequency of the received 
down-link signal is compared with a clock located at the reception station.
The Doppler observable is defined from the ratio of the emitted and 
received frequencies after correcting for the known constant factor.
The frequencies are measured against clocks which tick according to
proper times $d\sxm$ and $d\sxp$, so that the round trip Doppler shift 
$\ynu$ is built up on the ratio of these quantities
\beqa
\label{frequency_shifts}
e^{\ynu} &\equiv& {d \sxp \over d\sxm}
\eeqa
Although only the linear part will be needed,
this precise convention has been chosen for the sake of simplifying forthcoming 
expressions. As $d\sxm$ and $d\sxp$ obey (\ref{proper_times}), the round trip 
Doppler shift is finally given by
\beqa
\label{frequency_shift}
e^{\ynu} &=& {(\uxm\cdot \nxm) \over (\uy \cdot \nym)}
{(\uy\cdot \nyp) \over (\uxp\cdot \nxp)} 
\eeqa
This is a properly defined observable which is explicitly 
invariant under gauge changes. It must also be emphasized that it 
accounts for the effects of motions (Doppler effect) and 
of gravitation (Einstein redshift effect).

The Pioneer observable is then obtained by looking at the Doppler shift 
(\ref{frequency_shift}) as a function of time. Following \cite{Anderson02},
we convene to write it as an acceleration $\a$ with
\beqn
\label{shift_derivative}
2 \a d\s \equiv -\c^2 d \ynu
\eeqn
At this point, $d\s$ represents the proper time delivered by a reference
clock to be specified later on. Note that the definition of $\a$ thus depends 
on the choice of this reference clock whereas the definition of $\a d\s$ does not.
We also assume the observers and probe to follow geodesics, that is 
motions with vanishing covariant acceleration $\D\uxm = \D\uxp = \D\uy = 0$ 
(corrections associated with the position of stations on Earth
are taken into account separately).
Using the expression (\ref{frequency_shift}), we thus obtain the relation 
\beqa
\label{Doppler_acceleration}
{2\a d\s \over \c^2} &=& -{\left(\uxm \cdot \D\nxm \right) \over(\uxm \cdot\nxm)} +
{\left(\uxp \cdot \D\nxp \right) \over(\uxp \cdot\nxp)} \nonumber\\
&&-{\left(\uy \cdot \D\nyp \right) \over(\uy \cdot\nyp)} +
{\left(\uy \cdot \D\nym \right) \over(\uy\cdot \nym)} 
\eeqa
This relation is exact in the context of our simplifying assumptions. 
As already emphasized, the small corrections associated with perturbations
not taken into account in this context are supposed to be linearly 
superposed to the anomaly and, therefore, to disappear in the 
forthcoming expressions obtained after a subtraction.
 
The Doppler acceleration (\ref{Doppler_acceleration}) has a twofold dependence 
versus the metric. 
First, the Christoffel symbols which enter the covariant 
derivatives have their deviation from standard form given by 
(\ref{anomalous_acceleration_field}). 
Second, the time delays which enter the expressions 
of the anomalous wavevectors are deduced from (\ref{directions_waves}) as 
\beqa
\label{modified_wavevectors}
&&(\n^\pm_a)_\mu = \stand{(\n^\pm_a)_\mu} + (\delta\n^\pm_a)_\mu \quad,\quad
(\delta\n^\pm_a)_0 = 0\nonumber\\
&&(\delta\n^\pm_a)_i = \mp \c{\partial \tdan\over \partial (\xb_a^\pm)^i} 
\eeqa
$\tdan$ is the anomalous part of the time delay function and is 
computed in \ref{appendix_time_delay}.
One deduces the variations of the Doppler shift (\ref{frequency_shift})
\beqa
\label{time_delay_anomaly}
{d\sxp \over d\sxm} &=& 
{(\uxm \cdot \stand{\nxm}) \over (\uy \cdot \stand{\nym})} 
{(\uy\cdot \stand{\nyp}) \over (\uxp \cdot \stand{\nxp})} 
\left( 1 +  \ynuan \right) \nonumber\\
\ynuan &=& - {1\over (\uy \cdot \stand{\nym})}{\c d\tdan(\xbxm,\xby) \over d\sy} 
\nonumber\\
&&- {1\over (\uy \cdot \stand{\nyp})}{\c d\tdan(\xbxp,\xby) \over d\sy} 
\eeqa

Collecting these results, we obtain the anomalous acceleration $\aan$ 
by subtracting the value calculated in standard Einstein theory 
($\delta\PP=\PNan=0$) from the modified value calculated with $\delta\PP\neq 0$
or $\PNan \neq 0$. This quantity $\aan$ corresponds to the
Pioneer anomaly denoted by $\a_P$ in \cite{Anderson02}.
We write it as the sum of two contributions $\aant$ and $\aang$ corresponding
to the anomalies induced by time delay and Christoffel symbols respectively 
\beqa
\label{extra_acceleration}
\aan &=& \aant  + \aang \\
{2\aant d\sy\over\c^2} &=& -d \ynuan \nonumber\\
{2\aang d\sy\over\c^2} &=& d\sxm {(\delta \Gxx\cdot\nxm)\over(\uxm \cdot\nxm)}
- d\sxp {(\delta \Gpxx \cdot\nxp) \over (\uxp\cdot \nxp)}
\nonumber\\
&+& d\sy \left( {(\delta \Gyy \cdot\nyp) \over (\uy \cdot\nyp)}
- {(\delta \Gyy \cdot\nym) \over (\uy \cdot\nym)} \right)
\nonumber
\eeqa
We have now specified the reference clock to be on the probe ($d\s=d\sy$)
but another choice would not affect this expression at the first order
in the gravitation potentials. We have also replaced all quantities
but $\tdan$ and $\delta\Gamma$ by their zeroth order approximation.

Using equations (\ref{acceleration_field}), one sees that the contributions of the 
Christoffel symbols almost compensate in (\ref{extra_acceleration}) for the
Earth which has a nearly vanishing radial velocity.
Contributions from the Earth motion may then be neglected so that the anomalous acceleration 
(\ref{extra_acceleration}) may be rewritten, at leading order,
\beqa
\label{acceleration_anomaly_approximation}
{2\aant \over \c^2} &\simeq&  {d^2 \over d\sy^2}\left(c\tdan(\xbxm,\xby) + 
c\tdan(\xbxp,\xby)\right)\nonumber\\
{2\aang \over \c^2} &\simeq&  - 2 \delta \Gamma_{\ry,\sy\sy}
\eeqa 
Inserting in equations (\ref{acceleration_anomaly_approximation})
explicit expressions of time delays (\ref{perturbed_time_delay})
and Christoffel symbols (\ref{anomalous_acceleration_field})
while neglecting terms proportional to the angular 
velocity of the probe, one obtains the two contributions to the
anomaly in terms of the anomalous potentials $\PNan$ and $\delta\PP$
\beqa
\label{anomalous_acceleration}
{\aant \over \c^2} &\simeq& -{d \over d\sy} \left( 
(2 \PNan - \delta\PP){d\ry\over d\sy}\right) \nonumber\\
{\aang \over \c^2} &\simeq&   {d \delta \PN \over d\ry}
+ {d\delta\PP \over d\ry} \left({d\ry\over d\sy}\right)^2
\eeqa

It was shown in the preceding section that the anomalous part $\PNan$ of the 
first potential $\PN$ had to remain very small to fit planetary data.
This implies that the potential anomalous acceleration is mainly ascribed 
to the second potential $\PP$. We will see in the section \ref{sect:discussion}
that the terms $\aant$ and $\aang$ thus have similar contributions. Incidentally 
this will prove that none of the two contributions could be disregarded.

\section{Light deflection}
\label{sect:time_delay}

As argued in section \ref{sect:metric_tensors}, the linearized treatment 
is sufficient for probes having a large kinetic energy and this certainly
includes the study of light waves. 
We now study the effect of the second potential $\PP$
on Eddington-like light deflection experiments \cite{Will}, building up 
the description on the treatment of the time delay function and Doppler shifts 
presented in section \ref{sect:Pioneer}.

There are indeed close similarities between Doppler tracking of spacecrafts
and deflection experiments and the most recent deflection experiments 
use Doppler tracking techniques \cite{Bertotti03,Iess99}. 
As an important difference however, the effect of the time delay is favored 
by a kind of amplification during occultations by the gravitational source 
which play a key role in deflection experiments. 
As a consequence, the modifications of motions of endpoints are found to
play a negligible role in the observables. 
We show below that, in these conditions, the deflection of light is described 
by an effective Eddington $\gamma$ parameter, which is now a function 
rather than a constant. 
The usual expression of deflection can be used with the function $\gamma$ 
evaluated at the impact parameter $\ri$, that is the closest distance of 
approach of the light ray to the gravitational source. 
  
We consider a Doppler measurement on a down-link radio
signal from the probe to a station on Earth. 
We use the same notation as in the preceding section with $\uy$ and $\uxp$ 
denoting the space-time velocities of the emitter and receiver,
$\nyp$ and $\nxp$ the directions of wavevectors at emission and reception,
$d\sy$ and $d\sxp$ the respective proper times. We define the Doppler shift 
$\ynup$ on this one-way tracking technique by using (\ref{proper_times}) 
\beqn
e^{\ynup} \equiv {d\sxp \over d\sy} =
-{(\uy\cdot \nyp) \over (\uxp \cdot\nxp)}
\eeqn
As previously, this formula describes Doppler as well as Einstein effects.
For simplicity, we suppose the emitter to be at rest at a radius $\ry$ larger 
than the Sun-Earth distance $\rx=1$AU (with angle $\angy$) and the receiver on 
Earth to follow a circular orbit with constant radius $\rx$ (with angle $\angx$).
The angular velocity ${\dangx}$ of Earth is given by the third Kepler law
(see section \ref{sect:geodesics}). 
The Doppler shift is then deduced from the time delay function $\td$ 
\beqa
\label{one_way_Doppler}
&&e^{-\ynup} = K \left(1- {\d \td \over d\tx}\right) \\
&&K \equiv {\sqrt{\g_{00}(\ry)} \over 
\sqrt{\g_{00}(\rx) + \g_{\r\r}(\rx)\rx^2 \dangx^2/c^2}} 
\nonumber
\eeqa
$K$ is the ratio of rates for the emitter and receiver clocks. 
This factor was unity in the two-way technique studied in the preceding section 
since the two end-clocks were located on Earth.
With our simplifying assumptions, it is now a constant differing from unity.
Its constancy implies that the time delay function may be recovered 
by integrating the Doppler shift with respect to time.

Explicit expressions for the time delay are obtained in appendix 
\ref{appendix_time_delay}.
For propagation in a static and isotropic metric, the time delay only depends on the 
ratio $\g_{\r\r}/\g_{00}$, that is also in the linearized approximation 
on the gravitation potential $\Ps$. 
At first order, it is given by (\ref{perturbed_time_delay})
\beqa
\c \td &=& \rxy
-2 \int_\ri^\rx \Ps(\r) {\r d\r\over \sqrt{\r^2-\ri^2}} \nonumber\\
&&-2 \int_\ri^\ry \Ps(\r) {\r d\r\over \sqrt{\r^2-\ri^2}} 
\eeqa
$\rxy$ is the Minkowski spatial distance, 
i.e. the time delay in the absence of gravitational perturbation,
\beqa
\label{Minkowski_distance}
\rxy &=& \sqrt{\rx^2 - \ri^2} + \sqrt{\ry^2 -\ri^2}\nonumber\\
&=&\sqrt{\rx^2 +\ry^2 -2\rx\ry {\rm{cos}}(\angx-\angy)}
\eeqa
$\ri$ is the impact parameter and its time variation is given by 
(\ref{angle_impact_parameter})
\beqn
{d \ri \over d \tx} = - {\sqrt{\rx^2-\ri^2} \sqrt{\ry^2-\ri^2}\over \rxy } 
\dangx
\eeqn
$\dangx$ is the angular velocity of Earth on its orbit, 
that is also the cause of variation of $\ri$ in the deflection experiment.

We now discuss anomalies of the time delay, deflection angle or Doppler shift,
due to deviations of the potential $\Ps$ from its standard Einsteinian form.
To this aim, it is convenient to decompose $\Ps$ into a Newtonian part, isolating 
the singularity at $\r=0$, and a residual part expected to become significant 
at large distances
\beqn
\label{ps_decomposition}
\r \Ps(\r) \equiv -{\Gs \M \over \c^2} + {\M \over\c^2} \r^2 \cs(\r)
\eeqn
We consider light rays passing near the solar disk and assume
a smooth variation of $\cs(\r)$. 
Neglecting terms of the order or smaller than $(\ri/\rx)^2$ or  $(\ri/\ry)^2$,
we obtain in appendix \ref{appendix_time_delay} the following 
approximate expression for the anomalous time delay 
\beqa
\label{td_v_anomalies}
\c \tdan &\simeq&  \delta \gamma(\ri) {\GN\M\over\c^2}
{\rm{ln}}{4 \rx\ry\over \ri^2} \\
&-& {2\M\over\c^2}\left(\int_0^\rx \cs(\r) \r d\r 
+ \int_0^\ry  \cs(\r) \r d\r\right)\nonumber
\eeqa
where $\delta \gamma$ is the anomalous contribution to an effective  
Eddington parameter $\gamma$ 
\beqn
\label{gamma_anomaly}
\delta \gamma(\ri) \equiv {2(\Gs -\GN)  -\cs(\ri)\ri^2\over\GN}
\eeqn

We then deduce the anomalies for the deflection angle $\thnan$
and Doppler signal $\delta\ynup$
\beqa
\label{theta_y_anomalies}
\thnan &\simeq&  -{\ry \over \rx + \ry}{\partial (\c \tdan) \over \partial \ri}
\nonumber\\
\delta\ynup &\simeq&  {\partial (\tdan) \over \partial \tx}  \simeq
{\rx \dangx \over \c }\thnan 
\eeqa
Expressions (\ref{td_v_anomalies}) and (\ref{theta_y_anomalies}) contain the same 
information, as $\thnan$ corresponds to the derivative of $\tdan$ with respect to 
$\ri$ and $\delta\ynup$ to its derivative with respect to $\tx$. 
Note that the second term in expression (\ref{td_v_anomalies}) is constant and 
does not contribute to (\ref{theta_y_anomalies}). Both expressions 
remain valid when the emitter is far outside the solar system, provided its distance
$\ry$ be replaced by the cutoff distance $\rL$ (see the discussions at the end of 
section \ref{sect:gravitation_equations} and in the appendix). 

The usual PPN expressions are recovered when $\cs(\ri)$ vanishes, so that 
$\delta\gamma(\ri)$ reduces to a constant.
The time delay and deflection angle thus take simpler forms
\beqa
\label{ppn_anomalies}
\c \tdan_{\rm PPN} &\simeq&  \delta \gamma_0 {\GN\M\over\c^2} 
{\rm{ln}}{4 \rx\ry\over \ri^2} \nonumber\\
\thnan_{\rm PPN} &\simeq&  2 {\ry \over \rx + \ry}\delta \gamma_0 {\GN\M\over\c^2\ri}  
\eeqa
with $\delta\gamma_0$ identified as the anomalous part $(\gamma-1)$ of the 
Eddington parameter $\gamma$ 
\beqa
\delta \gamma_0 &\equiv& {2(\Gs -\GN)\over\GN} = -{\GP \over\GN} 
\eeqa
The main difference with the more general expressions (\ref{td_v_anomalies}) and 
(\ref{theta_y_anomalies}) is that the dependence of $\delta\gamma$ on $\ri$ 
potentially present in these expressions has disappeared from the PPN expressions
(\ref{ppn_anomalies}).
This points at a remarkable consequence of the extended framework, namely the fact
that a deflection test may contain more information than a constant anomaly
$\delta\gamma_0$. Indeed, a continuous measurement of the deflection angle during 
the occultation by the Sun may in principle provide one with the dependence 
(\ref{theta_y_anomalies}) of the deflection angle versus the impact parameter.

Using again the fact that the anomalous part $\PNan$ of the first potential $\PN$ 
has to remain very small, this entails that precise information on the second 
potential $\PP$ could become accessible from a confrontation to observations
of the formula (\ref{theta_y_anomalies}).
An observation of a variation of $\delta\gamma$ with $\ri$ would constitute
a clear confirmation of the extended framework proposed in this paper.
In any case, deflection data have to be confronted against Pioneer data since
they are determined by the same function $\PP$.
This point is discussed in a more quantitative manner in the next section.

\section{Discussion}
\label{sect:discussion}

We now collect the results obtained in the previous sections to confront
available observational or experimental data to the extended framework 
studied in this paper.
We first write down quantitatively the stringent bounds on the anomalous part 
$\PNan$ of $\PN$ which are drawn from planetary data.
We then focus the attention on the contribution of the second 
potential $\PP$, considering successively the problems of Pioneer-like 
anomalous accelerations and Eddington deflection experiments.
For the sake of making the discussion more precise, we occasionally use the 
simple model (\ref{Newton_Prime_constants}) as a benchmark,
keeping however in mind that the extended framework naturally accomodates a more
general spatial dependence of the gravitation potentials $\PN$ and $\PP$.

Constraints on $\PNan$ may be drawn from a comparison of the orbital frequencies 
of different planets, provided the radius of the orbit is measured through 
optical ranging, that is independently of the measurement of the orbital period
\cite{Talmadge88}. 
This has been done for example for Mars, 
using telemetry data from Viking probes \cite{Anderson96}.
Constraints on the anomaly of the Newton force $\partial_\r\PN$ are then 
deduced by using the third Kepler law (\ref{third_Kepler_law}).
This argument was already presented in section XI-B of \cite{Anderson02}. 
It was developed in \cite{Jaekel04b}, using the most recent bounds for
violations of Newton law at distances of the order of planetary orbital radii
\cite{Coy03}.

We write here the resulting bound as a difference of the ratios 
$\partial_\r\PNan / \partial_\r\Pst$ at the radius of Mars ($\rM \sim 1.5$AU)
and Earth ($\rT = 1$AU)
\beqn
\label{Kepler_bound}
\left\vert {\partial_\r\PNan(\rM)\over \partial_\r\Pst(\rM)}
- {\partial_\r\PNan(\rT)\over \partial_\r\Pst(\rT)} \right\vert \le 10^{-10}
\eeqn
In the simple model (\ref{Newton_Prime_constants}), this is read as a 
bound on the anomalous parameter $\cN$  
\begin{eqnarray}
\label{zetaN_bound}
\left\vert \cN \MS\right\vert  \le 10^{-10}\times
 \frac{\GN \MS}{\rM^{2}}
\simeq 5\times 10^{-13} \rm{m}\rm{s}^{-2} &&
\end{eqnarray}
This value is small enough to discard any possibility that the Pioneer anomaly 
could be explained from a long-range modification of the Newton law \cite{Anderson02}. 
In other words, an adhoc modification of first potential $\PN$ designed to fit the 
Pioneer anomaly would produce an effect on outer planets more than 1000 times
too large to remain unnoticed \cite{Jaekel04b}. 
Now, we have shown in the present paper that the second potential $\PP$ 
leaves the third Kepler law unaffected whereas it has an influence 
on Doppler tracking of probes with eccentric motions.
This opens free space for a gravitational interpretation of the Pioneer 
anomaly without conflict with planetary data \cite{Jaekel04a}.

From now on, we disregard the anomalous part $\PNan$ of the potential $\PN$ 
which has been shown to have a negligible effect.
We focus the attention on the contribution of the second potential $\PP$ to the 
Pioneer anomalous acceleration (\ref{anomalous_acceleration}) which is now read 
\beqa
\label{remote_extra_acceleration}
{\aan\over\c^2} &\simeq& 2 {d \PP \over d\ry} \left({d \ry \over d\sy}\right)^2
+ \PP {d^2 \ry \over d\sy^2}
\eeqa
The term proportional to the kinetic energy in (\ref{remote_extra_acceleration}) is 
found to dominate the other one which scales as the gravitational energy of the probe. 
Keeping only the former, we note that the anomalous acceleration is proportional to 
the kinetic energy of the probe, which is a remarkable prediction of the framework 
studied in the present paper.
Should we have at our disposal data points associated with probes with
very different kinetic energies, this prediction could be used to confirm or
infirm the framework.

In the simple model (\ref{Newton_Prime_constants}), the anomalous acceleration 
(\ref{remote_extra_acceleration}) is read as 
\beqn
\label{Pioneers_acceleration}
\aan \simeq  2\left(\cP \M + {\GP \M \over \dP^2}\right) {\vP^2 \over \c^2}
\eeqn
In this equation, the acceleration $\GP \M / \dP^2$ is much smaller than the 
Newton acceleration $\GN \M / \dP^2$ since the ratio 
$\vert\GP/\GN\vert\simeq\vert\delta\gamma_0\vert$ is much smaller than unity. 
It is also smaller than the acceleration $\cP\M$ if the latter is to be 
identified with the Pioneer anomaly (see below).
As the Pioneer 10/11 probes follow an escape trajectory towards the 
outer solar system with their velocities remaining practically constant 
\beqn
\vP \simeq 1.2 \times 10^{4} {\rm{m s}}^{-1} \quad, 
\quad {\vP^2 \over c^2} \simeq 1.6 \times 10^{-9}
\eeqn
the anomalous acceleration $\aan$ is found to be approximately 
constant when the distance to the Sun varies, in conformity with one of the most
striking properties of Pioneer observations \cite{Anderson02}.
Let us however emphasize that this conclusion has been obtained after various
simplifications recalled in the present section whereas the more general expression 
written above may in principle be confronted against more detailed data.

If we now make the jump of identifying the anomalous acceleration $\aan$ recorded on 
the Pioneer 10/11 probes with the effect (\ref{Pioneers_acceleration}) induced by the 
second potential $\PP$, we get an evaluation of the constant $\cP \M$
\beqa
\label{Pioneer_values}
\aan &\simeq&  2 \cP \M {\vP^2 \over \c^2} \sim 8 \times 10^{-10} \rm{m}\rm{s}^{-2}
\nonumber\\
&\rightarrow&  \quad \cP \M \sim 0.25 \rm{m}\rm{s}^{-2}
\eeqa
Note that the positive sign of $\aan$ is associated with a blue shift of Doppler
frequencies, with the same convention as in \cite{Anderson02}.
Hence, the identification (\ref{Pioneer_values}) entails that the constant 
$\cP$ is positive. 
Comparing with (\ref{zetaN_bound}), one also notes the important difference of magnitude
between the linear coefficients $\cN$ and $\cP$ in the two gravitation potentials.

At this point, it is worth coming back to the conditions of validity of the linearized 
approach which were discussed at the end of section \ref{sect:metric_tensors}.
Putting the numerical values obtained in (\ref{zetaN_bound}) and (\ref{Pioneer_values})
in these discussions leads to an a posteriori justification of the perturbative 
treatment performed at first order around Minkowski metric.
For escape trajectories towards the outer part of the solar system, second order terms
in the gravitation potentials are negligeable when compared with 
terms proportional to linearized potentials and kinetic energies.
This stands in contrast with the assumption of a comparable magnitude for kinetic and 
gravitational energies on which the PPN framework is built up. 
While such an assumption is well suited for describing the circular 
motion of planets, which justifies its key role in the PPN framework, 
it is also clear that it does not hold
for the discussion of Pioneer-like eccentric motions.

We now come to the discussion of deflection experiments which are
determined by the same potential $\PP(\r)$.
To this aim, we first rewrite the definition of the anomalous part $\delta\gamma$
of the Eddington parameter, using the fact that $\PNan$ may be disregarded.
We decompose $\PP$ as was done for $\Ps$ in (\ref{ps_decomposition}),
thus separating a part singular at $\r=0$ and a regular part describing 
the long range effect
\beqn
\r \PP(\r) \equiv -{\GP \M \over \c^2} + {\M \over\c^2} \r^2 \cP(\r)
\eeqn
As $\cN(\r)$ is negligible, $\cP(\r)$ is identical to $-2\cs(\r)$
with $\cs(\r)$ the quantity used in the preceding section.
For rays passing near the surface of the Sun, 
we then obtain the anomalous time delay (\ref{td_v_anomalies})
\beqa
\c \tdan &\simeq&  \delta \gamma(\ri) {\GN\M\over\c^2}
{\rm{ln}}{4 \rx\ry\over \ri^2} \nonumber\\
&+& {\M\over\c^2}\left(\int_0^\rx \cP(\r) \r d\r
+ \int_0^\ry  \cP(\r) \r d\r\right)\nonumber\\
\delta \gamma(\ri) &\equiv& \delta \gamma_0 +{\cP(\ri)\ri^2\over 2\GN}
\eeqa
We have assumed a smooth variation of $\cP(\r)$ around $\r=0$ and neglected 
terms which are not amplified near the occultation. We then rewrite
the anomaly (\ref{theta_y_anomalies}) of the deflection angle as
\beqn
\thnan \simeq  {\ry\over\rx+\ry}{\GN\M\over\c^2} \left( {2 \delta\gamma(\ri) \over \ri}  
- {\partial \delta\gamma(\ri) \over \partial \ri}  {\rm{ln}}{4 \rx\ry\over \ri^2} \right)
\eeqn
The PPN expressions are recovered when $\cP(\ri)$ vanishes, 
so that $\delta\gamma(\ri)$ reduces to the constant $\delta\gamma_0$. 

As already discussed, the specific feature of the extended framework is the  
dependence of $\delta\gamma$ on $\ri$ which reveals the long range variation of the 
potential $\PP$.
This specific feature can be looked for in deflection experiments by keeping trace
of the whole functional dependence of the anomalous deflection angle $\thnan$ 
during an occultation by the Sun 
\beqn
\label{Cassini_anomaly}
{\ri\thnan\over\rS\thnS} \simeq {\delta\gamma(\ri) \over 2}
- {\ri \partial \delta\gamma(\ri) \over 4 \partial \ri}  {\rm{ln}}{4 \rx\ry\over \ri^2} 
\eeqn
In order to define manifestly dimensionless quantities, we have introduced the Sun 
radius $\rS$ and the standard deflection angle $\thnS=4\GN\M/\c^2\rS$ for a ray grazing the Sun.
For most experiments, $\rx$ has to be replaced by the Sun-Earth distance $\dT$ 
and $\ry$ by the Sun-probe distance. 
When the emitter is a source far outside the solar system,
$\ry$ has to be replaced by the cutoff distance $\rL$.
Should the precision of the measurement be sufficient, we could reconstruct the 
$\ri$-dependence of $\delta\gamma$ and, then, the $\r$-dependence of $\cP$ and $\PP$.

We want to stress that the purpose of extracting the 
$\ri$-dependence of the effective parameter $\delta\gamma$
is different from that of existing analysis which aim at obtaining a global value 
of $\delta\gamma$ on the whole range of available values of $\ri$.
Formula (\ref{Cassini_anomaly}) suggests that deviations should tend to vary with the impact parameter,
being larger when the impact parameter increases, that is also when the deflection angle decreases,
so that the detection strategy cannot be the same as that used in existing analysis.
Consequently, the constraints drawn for example from Cassini \cite{Bertotti03} 
or VLBI measurements \cite{Shapiro04} cannot easily be translated 
into a constraint on the $\ri$-dependent contribution to $\delta\gamma$.
We can only make a guess whether or not
these measurements can produce information of interest for our purpose
by proceeding as follows.
 
If we consider the simplest model (\ref{Newton_Prime_constants}) with $\cP$ 
constant over the whole range of distances from the radius of the Sun to the 
size of the solar system, we may use the value (\ref{Pioneer_values}) of $\cP$ 
deduced from Pioneer data as a first guess for the Cassini experiment. 
We thus obtain from (\ref{Cassini_anomaly})
\beqn
\label{Cassini_variation}
{\ri\thnan\over\rS\thnS} \simeq {\delta\gamma_0 \over 2}
- {\cP\M\ri^2\over\c^2\rS\thnS} {\rm{ln}}{4 \rx\ry\over e\ri^2} 
\eeqn
This quantity varies by a few $10^{-3}$ when $\ri$ goes from 5 to 10 $\rS$, which 
should probably be visible in Cassini data \cite{Bertotti03}, if looked for.
Emphasizing once again that the signature we are interested in is a potential 
variation of $\delta\gamma$ with $\ri$ rather than a mean value of $\delta\gamma$,
this result certainly pleads for a reanalysis of Cassini data \cite{Bertotti03} 
aimed at the observation of such a variation. 
Such an observation would constitute a clear indication for a non null value 
of $\cP$, to be then confronted to the value deduced from Pioneer data.  
Different values could also be obtained in the two experiments 
since $\cP$ is in fact a function of $\r$ which could vary 
when $\r$ goes from the radius of the Sun $\rS\sim 0.7\times 10^{9}$m 
to the distance explored by the Pioneer probes $\dP\sim 1.2\times 10^{13}$m.

\section{Conclusion}

We have shown that Einstein gravitation theory possesses natural extensions 
characterized by two running gravitation couplings replacing the single Newton 
constant or, equivalently, two gravitational potentials $\PN$ and $\PP$
replacing the usual Newton potential $\Pst$. 
The running couplings may depend on the length scale 
and differ in the two sectors of traceless and traced tensors, 
which have different conformal weights.  

The good agreement between planetary data and General Relativity
essentially means that the long range modification of the first potential
$\PN$ has a negligible effect. This conclusion has been drawn from the discussion of 
third Kepler law for circular orbits. It would also be worth discussing the
perihelion precession but this study, which involves nonlinearities of the gravitation 
theory, is postponed to a forthcoming analysis.
The linearized treatment performed in the present paper encompasses the phenomena 
for which the kinetic energy is much larger than the gravitational one,
that is also those usually ascribed to the $\gamma$ parameter in the PPN formalism.
This includes the case of spacecrafts following escape trajectories towards the
outer solar system as well as deflection experiments on light rays.

The second potential $\PP$ has been shown to have a crucial effect on 
Doppler tracking of probes with highly eccentric motions,
allowing one to propose a gravitational interpretation of the Pioneer anomaly
without raising a conflict with planetary tests. 
This interpretation comes with the crucial prediction of 
a magnitude of the anomalous acceleration proportional to the 
kinetic energy of the probe. 
This prediction cannot be confronted against the data available today for the 
Pioneer 10/11 probes which were endowed with very close kinetic energies, but 
it could be the subject of dedicated analysis of Doppler tracking data
stored during their flight to the outer solar system. In particular, a 
study of the fly-by sequences of the probes might provide interesting clues.

While the presence of $\PP$ at scales of the order of the solar system size 
is tested by Pioneer-like Doppler tracking, its variation in the vicinity
of the solar radius can be studied through deflection experiments. 
We have shown that $\PP$ may thus be seen as promoting the anomalous 
Eddington parameter $(\gamma-1)$ to the status of a function 
of the impact parameter $\ri$. 
Evidence of a variation of $\gamma$ would provide a direct
validation of the extended framework. 
This aim might be attainable through a reanalysis of existing data,
in particular those of the Cassini experiment \cite{Bertotti03}.
On a longer term, it can also be reached by high accuracy Eddington tests (see
for example the LATOR project \cite{Lator}) or global mapping of deflection
over the sky (see for example the GAIA project \cite{Gaia}). 
The former would lead to an accuracy on the measurement of $\gamma$ certainly
sufficient to see the effect if its order of magnitude is given by the numbers
deduced from the Pioneer data in the preceding section. Meanwhile the latter
would allow one to reconstruct the dependence of $\gamma$ versus the impact
parameter $\ri$ and then, the dependence of $\PP$ versus $\r$. 
Should these future projects conclude to the absence of such a dependence,
this would anyway result in improved constraints on a family
of natural metric extensions of General Relativity, confirming
its validity for describing gravity in the solar system.   

\appendix
\section{Time delay function}
\label{appendix_time_delay}

In this appendix, we determine explicit expressions for the time delay function $\td$ 
which characterizes light propagation. $\td$ is a two-point function derived from the 
phase, that is also, in the semi-classical limit, the action $\S$ which satisfies 
Hamilton-Jacobi equation. 

Stationarity and isotropy of the metric entail the conservation laws (\ref{conservations}). 
It follows that the action $\S$ can be written in terms of a 
phase shift $\Sr$ which is a function of $\r$ only (see (\ref{phase_shift})).
The phase shift obeys an Hamilton-Jacobi equation (\ref{Hamilton_Jacobi}) rewritten here
for light propagation (massless field)
\beqn
(\partial_\r \Sr)^2 = - {\g_{\r\r}\over \g_{00}}{\E^2\over\c^2} -
{\J^2 \over \r^2}
\eeqn
It can be given the following integral form determined by the conformally invariant 
ratio $\g_{\r\r}/\g_{00}$ of the metric components 
\beqn
\label{HJ_phase}
\Sr = \pm \J \int^\r d\r 
\sqrt{-{\E^2\over\J^2\c^2}{\g_{\r\r}\over\g_{00}} - {1\over \r^2}}
\eeqn
The same solution is equivalently written as a characteristic equation for 
the associated null geodesics
\beqa
\label{lightlike_trajectory}
&&{d\ang \over d\r} = 
{\pm 1/\r^2 \over 
\sqrt{-{\E^2\over\J^2\c^2}{\g_{\r\r}\over\g_{00}} - {1\over \r^2}}}
= - \partial_\J \partial_\r \Sr
\eeqa

A null geodesic is essentially characterized by the 
radial coordinate $\ri$ at closest approach of the Sun.
Equation (\ref{HJ_phase}) is then solved in terms of the reduced function 
\beqn
\wt(\x) \equiv {\g_{00} \over \g_{rr}} (\ri) {\g_{rr} \over \g_{00}} (\r) 
\quad,\quad \x \equiv {\ri\over\r}
\eeqn 
Fixing by convention the value of $\Sr$ to be zero at closest approach,
its explicit expression is obtained as 
\beqa
\label{null_geodesic_solutions}
&&\Sr(\r) = - \J \sw \left({\ri \over \r}\right) \quad,\quad \ang \le \angi\nonumber\\
&&\Sr(\r) = + \J \sw \left({\ri \over \r}\right) \quad,\quad \ang \ge \angi\nonumber\\
&&\sw(\x) \equiv \int_\x^1  {d\x \over \x^2} \sqrt{\wt(\x) - \x^2}
\eeqa 
The geometry of the null geodesic is then determined by writing boundary
conditions on its enpoints 
\beqa
&&\ang(\r) + \cw \left({\ri \over \r}\right) = \angi,
\qquad \ang \le \angi\nonumber\\
&&\ang(\r) - \cw \left({\ri \over \r}\right) = \angi,
\qquad \ang \ge \angi\nonumber\\
&&\cw(\x) \equiv -\partial_\J \left( \J \sw(\x) \right)
= \int^1_\x  {d\x \over \sqrt{\wt(\x) - \x^2}}
\nonumber\\
\label{boundary_conditions}
&&\angy + \cw \left({\ri \over \r}\right) = \angi =
\angx - \cw \left({\ri \over \r}\right)  
\eeqa
Note that equations (\ref{boundary_conditions}) simultaneously determine 
the distance of closest approach and the angles $\angi$ 
as functions of the endpoints.

The time delay, that is the elapsed time during propapagation of a constant
phase from $(\rx, \angx)$ to $(\ry, \angy)$, is then deduced from the 
phase shift (\ref{null_geodesic_solutions})
\beqa
\label{phase_shift_time_delay}
\tdxy &\equiv& \td(\rx,\angx,\ry,\angy) \\
&=& {1 \over \E}\left( \J(\angx - \angy)
+ \Sr(\rx) - \Sr(\ry)\right) \nonumber\\
&=& \sqrt{-{\g_{\r\r}\over \g_{00}}(\ri)}{\ri\over \c}\left( \angx - \angy
+ \sw \left({\ri \over \rx}\right) + \sw \left({\ri \over \ry}\right)  \right)
\nonumber
\eeqa
The wavevector, tangent to the null geodesic, can be obtained from the endpoint 
variations of the time delay
\beqa
\label{time_delay_variations}
d_1(\c \tdxy) &=& -\sqrt{-{\g_{00}(\rx)\over \g_{\r\r}(\rx)}} \\
&&\times\left( \cos (\thnx -\angx)d\rx 
+ \sin (\thnx -\angx) \rx d\angx\right)\nonumber
\eeqa
Here $d_1$ represents a variation of the endpoint $(\rx,\angx)$ with the other one
kept fixed and $\thnx$ denotes the line of sight angle, that is the angle 
associated with the light propagation direction.
Variations with respect to the second endpoint may be recovered by symmetry. 
Equations written up to now hold at any order in the gravitational perturbation.

In the absence of gravitational fields, $\c \tdxy$ is given by the spatial distance 
$\rxy$ evaluated in Minkowski metric (see eq.(\ref{Minkowski_distance})).
Meanwhile null geodesics are straight lines  
and the line of sight angle is preserved under propagation
\beqn
\thny = \thnx = \thn_{(0)}
\eeqn
We now simplify the expression (\ref{phase_shift_time_delay}) of the time delay 
function up to first order in the potential $\Ps$ 
\beqa
\label{perturbed_time_delay}
\c \tdxy &=& \rxy \\
&-&2 \int_\ri^{\rx} \Ps(\r) {\r d\r\over \sqrt{\r^2-\ri^2}}
-2 \int_\ri^{\ry} \Ps(\r) {\r d\r\over \sqrt{\r^2-\ri^2}}\nonumber
\eeqa
Expanding relation  (\ref{time_delay_variations})
up to first order in $\Ps$, one obtains
the deflection angle $\Delta \thn$ at the same order
\beqa
\label{time_delay_variations2}
\Delta \thn &=& \thnx - \thn_{(0)} \\
d_1(\c \tdxy) &=& (1-2\Ps(\rx)) d\rxy +\rxy \Delta \thn d\thn_{(0)} \nonumber
\eeqa
Note that the derivative of the time delay along the line of sight is simply 
related to the gravitation potential $\Ps$.
Equation (\ref{time_delay_variations2}) involves the unperturbed angle $\thn_{(0)}$,
or equivalently the impact parameter $\ri$,  at lowest order 
\beqa
\label{angle_impact_parameter}
d_1 \ri &=& - \sqrt{\ry^2 -\ri^2} d\thn_{(0)} \\
&=&  \sqrt{\rx^2 -\ri^2} (d\thn_{(0)}  - d\angx) + {\ri \over\rx}d\rx\nonumber
\eeqa
This entails that the deflection angle is simply related to the derivative of the
time delay  with respect to the impact parameter, at constant radial coordinate
$\rx$
\beqa
\label{defection_angle}
\Delta \thn &=&-{\sqrt{\ry^2 -\ri^2}\over \rxy}
{\partial (\c\td)\over\partial \ri} \nonumber\\
&&- (1 -2 \Ps(\rx)){\ri\over \sqrt{\rx^2-\ri^2}}
\eeqa

Expression (\ref{perturbed_time_delay}) shows amplification 
of the time delay near occultation.
In order to isolate the singularity at vanishing impact parameter
we decompose the potential $\Ps$ into its Newtonian singular part
and a residual representing the long distance modification,
according to (\ref{ps_decomposition}). We then rewrite it as
\beqa
\label{potential_decomposition}
\r \Ps(\r) &=& -\left(\Gs -{\cs(\ri)\ri^2\over 2}\right) {\M \over\c^2}
+\left(2\r^2 -\ri^2\right){\cs(\ri)\over2}{\M \over\c^2}\nonumber\\
&+& \r^2 \left(\cs(\r) -\cs(\ri)\right) {\M \over\c^2}
\eeqa
In order to allow for a simple comparison with the PPN framework, we now 
introduce a function $\gamma(\ri)$ which generalizes the Eddington parameter
(see eq.(\ref{gamma_anomaly}))
\beqn
1+\gamma(\ri) \equiv {2\Gs  -\cs(\ri)\ri^2\over\GN}
\eeqn
The time delay (\ref{perturbed_time_delay}) then takes the explicit form 
\beqa
\label{time_delay_perturbation}
\c\tdxy &=& \rxy \\
&-& \left( 1+\gamma(\ri)\right) {\GN\M\over\c^2} {\rm{ln}}
{\rx - \rxy + \ry \over \rx + \rxy + \ry} \nonumber\\
&-&{\M\over\c^2}\left((2\cs(\rx) - \cs(\ri))\rx\sqrt{\rx^2-\ri^2} \right.\nonumber\\ 
&&\,\left. + \left(2\cs(\ry) - \cs(\ri)\right)\ry\sqrt{\ry^2-\ri^2} \right)\nonumber\\
&+&{2\M\over\c^2} \left( \int^{\rx}_\ri \sqrt{\r^2-\ri^2}
d\left(\r(\cs(\r)-\cs(\ri))\right) \right.\nonumber\\ 
&&\,\left. + \int^{\ry}_\ri \sqrt{\r^2-\ri^2}d\left(\r(\cs(\r)-\cs(\ri))\right)\right)
\nonumber
\eeqa
The logarithmic part has the same form as in the PPN formalism, but with $\gamma$ 
now a function of the impact parameter $\ri$. 
This dependence is determined by the long range behavior of the potential $\Ps$, 
that is also by the function $\cs$. Assuming that $\cs$ is a regular function,
the last terms of equation (\ref{time_delay_perturbation}) may be expanded
in $\ri/\rx$ and $\ri/\ry$, so that the time delay 
(\ref{perturbed_time_delay}) may be rewritten, up to terms 
equal to or smaller than  $(\ri/\rx)^2$ or $(\ri/\ry)^2$
\beqa
\label{pert_time_delay}
\c\tdxy &=& \rxy \\
&-& \left( 1+\gamma(\ri)\right) {\GN\M\over\c^2} 
{\rm{ln}}{\rx - \rxy + \ry\over \rx + \rxy + \ry} \nonumber\\
&-& {2\M\over\c^2}\left(\int_0^\rx \cs(\r) \r d\r
+ \int_0^\ry  \cs(\r) \r d\r\right)\nonumber
\eeqa
In the same context, the deflection angle (\ref{defection_angle})
may be rewritten up to terms of order $\ri/\rx$ or $\ri/\ry$
\beqa
\label{remote_deviation_angle}
\Delta \thn &\simeq& -{\ry \over \rxy} {\partial (\c \td)\over \partial \ri}
\\
&\simeq& -{\ry \over \rx + \ry} {\GN\M\over\c^2}{\partial \over \partial \ri} 
\left(\left( 1+\gamma(\ri)\right) {\rm{ln}}{4\rx\ry\over\ri^2}\right)\nonumber
\eeqa
The logarithmic divergence with respect to $\ry$ 
in equation (\ref{remote_deviation_angle}) signals 
a failure of the perturbation expansion leading to equations 
(\ref{perturbed_time_delay}) and (\ref{time_delay_variations2}), due to the 
fact that the gravitational potential cannot be considered as weak
over large distances. To fix this problem, one must come back to 
exact expressions given previously for the deflection angle.
As discussed in section (\ref{sect:gravitation_equations}), 
the linear behaviour of the long range 
contribution to $\Ps$ can only be an approximation within the solar system  
of the true potential $\Ps$ which should remain weak over much larger distances. 
This behavior is thus limited to a maximal distance $\rL$, beyond which the 
potential recovers a purely Newtonian form. 
For instance, the cut-off may take the form of a mass parameter in a Yukawa 
potential with a range $\rL$ larger than the size of the solar system
\cite{Jaekel04b}.
The effect of the anomalous part of $\Ps$ from this range $\rL$ to very remote 
light sources may then be neglected, and the deflection angle 
obtained by replacing $\ry$ by $\rL$ in (\ref{remote_deviation_angle})
\beqa
\label{remote_deviation_angle2}
\Delta \thn \simeq - {\GN\M\over\c^2}{\partial \over \partial \ri}
\left(\left( 1+\gamma(\ri)\right) {\rm{ln}}{4\rx\rL\over\ri^2}\right)&&
\eeqa 

\def\etal{\textit{et al }}
\def\ibid{\textit{ibidem }}



\begin{thebibliography}{99}
\bibitem{Will} Will C. M., Theory and experiment in gravitational physics 
(1993) Cambridge U. P.; Living Rev. Rel. \textbf{4} (2001) 4.
\bibitem{Fischbach}  Fischbach E. and Talmadge C., 
The Search for Non Newtonian Gravity (1998) Springer Verlag. 
\bibitem{Aguirre}  Aguirre A., Burgess C.P., Friedland A. and Nolte D.,
{\it Class. Quant. Grav.} \textbf{18} (2001) R223.
\bibitem{Riess}  Riess A. G., Filippenko A. V., Challis P. \etal,
{\it Astron. J.} \textbf{116} (1998) 1009.
\bibitem{Perlmutter}  Perlmutter S., Aldering G., Goldhaber G. \etal, 
{\it Astrophys. J.} \textbf{517} (1999) 565;
Perlmutter S., Turner M. S. and White M.,
\textit{Phys. Rev. Lett.} \textbf{83} (1999) 670.
\bibitem{Sanders02} Sanders R. H. and McGaugh S. S., 
{\it Annu. Rev. Astron. Astrophys.} \textbf{40} (2002) 263.
\bibitem{Lue04}  Lue A., Scoccimari R. and Starkman G., 
{\it Phys. Rev.} {\bf D 69} (2004) 044005.
\bibitem{Turner04}  Carroll S.M., Duvvuri V., Trodden M. and Turner M.S., 
{\it Phys. Rev.} {\bf D 70} (2004) 043528.
\bibitem{Anderson98} Anderson J.D., Laing P.A., Lau E.L. \etal, 
{\it Phys. Rev. Lett.} {\bf 81} (1998) 2858.
\bibitem{Anderson02} Anderson J.D., Laing P.A., Lau E.L. \etal, 
{\it Phys. Rev.} {\bf D 65} (2002) 082004.
\bibitem{Anderson03} Anderson J.D., Lau E.L., Turyshev S.G. \etal, 
\textit{Mod. Phys. Lett.} \textbf{A17} (2003) 875.
\bibitem{Nieto04} Nieto M.M. and Turyshev S.G., \textit{Class. Quantum Grav.} 
\textbf{21} (2004) 4005.
\bibitem{Turyshev04} Turyshev S.G., Nieto M.M. and Anderson J.D.,
{\it 35th COSPAR Scientific Assembly} (2004) \eprint{gr-qc/0409117}.
\bibitem{Bertolami04} 
Bertolami O. and Paramos J., \textit{Class. Quantum Grav.} \textbf{21} (2004) 3309;
see also  \eprint{astro-ph/0408216} and  \eprint{gr-qc/0411020}.
\bibitem{deWitt} Utiyama R. and De Witt B., {\it J. Math. Phys.} 
{\bf 3} (1962) 608. 
\bibitem{Deser74}  Deser S. and van Nieuwenhuizen P., 
{\it Phys. Rev.} {\bf D 10} (1974) 401.
\bibitem{Capper74}  Capper D.M., Duff M.J. and Halpern L., 
{\it Phys. Rev.} {\bf D 10} (1974) 461.
\bibitem{Stelle}  Stelle K.S., {\it Phys. Rev.} {\bf D 16} (1977) 953;
{\it Gen. Rel. Grav.} {\bf 9} (1978) 353.
\bibitem{Sakharov} Sakharov A. D., {\it Doklady Akad. Nauk SSSR} 
{\bf 177} (1967) 70 [{\it Sov. Phys. Doklady} {\bf 12} 1040]. 
\bibitem{Adler}  Adler R.J., {\it Rev. Modern Phys.} {\bf 54} (1982) 729.
\bibitem{Nieto} Goldman T., P\'erez-Mercader J., Cooper F. and Nieto M. M.,
{\it Physics Lett.} {\bf B 281} 219.
\bibitem{Deffayet02} Deffayet C., Dvali G., Gabadadze G. and Vainshtein A., 
{\it Phys. Rev.} {\bf D 65} (2002) 044026.
\bibitem{Dvali03} Dvali G., Gruzinov A., Zaldarriaga M., 
{\it Phys. Rev.} {\bf D 68} (2003) 024012.
\bibitem{Gabadadze04} Gabadadze G. and Shifman M., 
{\it Phys. Rev.} {\bf D 69} (2004) 124032.
\bibitem{tHooft} t'Hooft G. and Veltman M., {\it Ann. Inst. H. Poincar\'e} 
{\bf A 20} (1974) 69.
\bibitem{Fradkin} Fradkin E. S. and Tseytlin A. A., {\it Nucl. Phys.} {\bf B 201}
(1982) 469.
\bibitem{Reuter} Reuter M. and Weyer H., \eprint{hep-th/0410119}.
\bibitem{Simon90} Simon J. Z., {\it Phys. Rev.} {\bf D 41}
(1990) 3720.
\bibitem{Hawking02} Hawking S. W. and Hertog T., {\it Phys. Rev.} {\bf D 65}
(2002) 103515.
\bibitem{DeFilippo02} De Filippo S. and Maimone F., 
{\it Phys. Rev.} {\bf D 66} (2002) 044018.
\bibitem{AndersonP03} Anderson P. R., Molina-Paris C. and Mottola E., 
{\it Phys. Rev.} {\bf D 67} (2003) 024026.
\bibitem{Hu04} Hu B. L., Roura A. and Verdaguer E., 
{\it Phys. Rev.} {\bf D 70} (2004) 044002.
\bibitem{Jaekel04a}  Jaekel M. T. and Reynaud S., ``Gravity tests in the solar system
and the Pioneer anomaly'', to be published in {\it Modern Physics Letters} {\bf A} (2005),
 \eprint{gr-qc/0410148}.
\bibitem{Shapiro66}  Shapiro I. I., {\it Phys. Rev.} {\bf 145} (1966) 1005.
\bibitem{Shapiro71}  Shapiro I. I., Ash M. E., Ingalls R. P. \etal, {\it Phys. Rev. Lett.} {\bf 26} (1971) 1132.
\bibitem{Reasenberg79} Reasenberg R. D., Shapiro I. I., MacNeil P. E. \etal, 
\textit{Astrophys. J.} \textbf{234} (1979) L219.
\bibitem{Shapiro99}  Shapiro I. I., {\it Rev. Mod. Phys.} {\bf 71} (1999) S41.
\bibitem{Bertotti03} Bertotti B., Iess L. and Tortora P., {\it Nature} {\bf 425}
(2003) 374.
\bibitem{Shapiro04}  Shapiro S. S., Davis J. L., Lebach D. E. and Gregory J. S., 
{\it Phys. Rev. Lett.} {\bf 92} (2004) 121101.
\bibitem{Adelberger03} Adelberger E. G., Heckel B. R. and Nelson A. E., \textit{%
Ann. Rev. Nucl. Part. Sci.} \textbf{53} (2003) 77, also in \eprint{hep-ph/0307284}.
\bibitem{Williams96} Williams J. G., Newhall X. X. and Dickey J. O., \textit{%
Phys. Rev.} \textbf{D53} (1996) 6730.
\bibitem{Hellings83} Hellings R. W., Adams P. J., Anderson J. D. \textit{et al}, 
\textit{Phys. Rev. Lett.} \textbf{51} (1983) 1609.
\bibitem{Anderson96} Anderson J. D., Gross M., Nordtvedt K. L. and Turyshev S. G., 
\textit{Astrophys. J.} \textbf{459} (1996) 365.
\bibitem{Damour} Damour T., {\it Class. Quantum Grav.} {\bf 13} (1996) A33;
Damour T., Piazza F. and Veneziano G., {\it Phys. Rev.} {\bf D 66} (2002) 046007.
\bibitem{Overduin00} Overduin J. M., {\it Phys. Rev.} {\bf D 62} (2000) 102001.
\bibitem{Landau} Landau L. D. and Lifschitz E. M., The classical theory of fields
(1975) Butterworth.
\bibitem{Jaekel95} Jaekel M. T. and Reynaud S., \textit{Annalen der Physik}
\textbf{4} (1995) 68.
\bibitem{Jaekel04b}  Jaekel M. T. and Reynaud S., ``Testing the Newton law at long 
distances'', to be published in {\it Int. J. Mod. Phys.} {\bf A} (2005),
 \eprint{gr-qc/0501038}.
\bibitem{Eddington} Eddington A. S., The mathematical theory of relativity (1957)
Cambridge U. P..
\bibitem{Robertson} Robertson H. P. in Space age astronomy (1962) Academic Press.
\bibitem{Schiff66} Ross D. H. and Schiff L. I., {\it Phys. Rev.} {\bf 141} (1966) 1215.
\bibitem{Nordtvedt68} Nordtvedt K., {\it Phys. Rev.} {\bf 169} (1968) 1014 \& 1017. 
\bibitem{WillNordtvedt72} Will C. M. and Nordtvedt K., 
{\it Astrophys. J.} {\bf 177} (1972) 757; Nordtvedt K. and Will C. M., \ibid 775.
\bibitem{Shapiro89} Shapiro I. I., in {\it General Relativity and Gravitation 1989} 
N. Ashby, D. F. Bartlett and W. Wyss eds (Cambridge UP, 1990) 313. 
\bibitem{effetNordtvedt} Nordtvedt K., {\it Phys. Rev.} {\bf 170} (1968) 1186. 
\bibitem{Iess99} Iess L., Giampieri G., Anderson J.D. and Bertotti B., 
{\it Class. Quant. Grav.} {\bf 16} (1999) 1487.
\bibitem{Talmadge88} Talmadge C., Berthias J.-P., Hellings R.W. and Standish E.M.,
 \textit{Phys. Rev. Lett.} \textbf{61} (1988) 1159.
\bibitem{Coy03} Coy J., Fischbach E., Hellings R.W., Talmadge C. and Standish E.M., 
private communication (2003).
\bibitem{Lator} Turyshev S.G., Shao M. and Nordtvedt K. Jr,
{\it Class. Quantum Grav.} {\bf 21} (2004) 2773.
\bibitem{Gaia} Vecchiato A., Lattanzi M.G., Bucciarelli B., 
{\it Astron. Astrophys.} {\bf 399} (2003) 337.
\end{thebibliography}
\end{document}